\documentclass[journal,a4paper,10pt]{IEEEtran}

\usepackage{tikz} 
\usetikzlibrary{shapes}
\usepackage{pgfplots}
\pgfplotsset{compat=1.5}
\usepackage[utf8]{inputenc} 
\usepackage[T1]{fontenc}
\usepackage{url}
\usepackage{ifthen}
\usepackage[noadjust]{cite}
\usepackage[cmex10]{amsmath}
\usepackage{multirow}
\usepackage{cite}
\usepackage{amsmath,amssymb,amsfonts}
\usepackage{algorithm}
\usepackage{subcaption}
\usepackage{algpseudocode}
\usepackage{graphicx}
\usepackage{makecell}
\usepackage{textcomp}
\usepackage{xcolor}
\usepackage{caption}
\usepackage{mathtools}
\usepackage{comment}
\usepackage{amsthm}
\usepackage{csquotes}
\usepackage{enumitem}
\usepackage{xcolor, soul}
\sethlcolor{yellow}
\usepackage{arydshln}
\usepackage{stfloats}
\usepackage{epsfig}
\newtheorem{thm}{Theorem}
\newtheorem{lem}{Lemma}
\newtheorem{prop}{Proposition}
\newtheorem{corollary}{Corollary}
\newtheorem{rem}{Remark}

\newtheorem{defn}{Definition}

\newtheorem{exmp}{Example}

\DeclarePairedDelimiter\ceil{\lceil}{\rceil}
\DeclarePairedDelimiter\floor{\lfloor}{\rfloor}

\newcommand\Osq{\mathbin{\text{\scalebox{.84}{$\square$}}}}

\begin{document}
\title{On Function-Correcting Codes}

\author{Rohit Premlal, 
	\IEEEmembership{Graduate Student Member, IEEE} and B. Sundar Rajan, \IEEEmembership{Life Fellow, IEEE}
	\thanks{This work was supported partly by the Science and Engineering Research Board (SERB) of Department of Science and Technology (DST), Government of India, through J.C. Bose National Fellowship to Prof. B. Sundar Rajan. Part of the content of this manuscript appeared in \textit{Proc. IEEE Inf. Theory Workshop (ITW)}, Nov. 2024, doi: 10.1109/ITW61385.2024.10807007. \cite{FCC_ITW}.	
			
		Rohit Premlal is with the Department of Electrical and Computer Engineering, University of Maryland, College
		Park, MD 20742 USA, and B. Sundar Rajan is with the Department of Electrical Communication Engineering, IISc Bangalore, India (e-mail: rohitpl@umd.edu, bsrajan@iisc.ac.in).  }
}

\maketitle

\begin{abstract}
Function-correcting codes were introduced in the work "Function-Correcting Codes" (FCC) by Lenz et al. 2023, which provides a graphical representation for the problem of constructing function-correcting codes. We use this function dependent graph to get a lower bound  on the redundancy required for function correction codes. Considering the function to be a bijection, leads to a lower bound on the redundancy required for classical systematic error correcting codes (ECCs). We propose a range of parameters for which this bound is tight. For single error correcting codes, we show that this bound is at least as good as a bound proposed by Zinoviev,
Litsyn, and Laihonen in 1998. Thus, this framework helps to study classical systematic error correcting codes.  Further, we study the structure of this function dependent graph for linear functions, which leads to bounds on the redundancy of linear-function correcting codes. We show that the Plotkin-like bound for function-correcting codes proposed by Lenz et.al 2023 is simplified for linear functions.  We identify a class of linear functions for which an upper bound proposed by Lenz et al., is tight and also identify a class of functions for which coset-wise coding is equivalent to a lower dimensional classical error correction problem.
\end{abstract}
\IEEEpeerreviewmaketitle
\section{Introduction}
\label{intro}
In a typical communication system, the sender wishes to transmit a message to the receiver through an erroneous channel. All symbols of the message are considered equally important and the objective is to create an error correcting code (ECC) with a decoder that can recover the entire message accurately. However, in certain scenarios, the receiver may be interested in a specific function of the message as well. In such cases, if the sender knows the function, the message can be encoded so as to ensure that the desired attribute (specific function of the message symbols) is protected against errors. This approach gives rise to a new class of codes known as function-correcting codes (FCCs) as introduced in \cite{FCC}.

The paradigm of FCCs as introduced in \cite{FCC}, looks at systematic codes that protect a general function evaluation of the message from errors. The channel considered is the substitution channel, in which the errors are symbol substitutions on the transmitted vector. The authors proposed a function-dependent graph for the problem of constructing FCCs. The vertex set of the graph represents possible codewords of the FCC and any two
vertices are connected if and only if they cannot both be contained together in an FCC. Thus, independent sets of the graph, of sufficient cardinality form FCCs. It is difficult to characterise the independent sets of this graph for a general function. An equivalence between FCCs and irregular-distance codes is shown in \cite{FCC}, where the distance requirements of the irregular-distance code is dictated by the function of interest. Since the problem heavily depends on the function, simplified sub optimal bounds on redundancy were proposed that are easier to evaluate. A fundamental result from classical coding theory is the Singleton bound, which states that to correct $t$ errors, at least $2t$ redundant symbols have to be added. It is shown in \cite{FCC} that this lower bound on redundancy holds for FCCs as well. It is also known that there are no non-trivial binary classical ECCs that achieve this lower bound. But in the case of FCCs, a class of functions called locally-binary functions have been proposed for which there exists encoding and decoding schemes so that $t$ errors can be corrected using $2t$ redundant symbols, for a particular range of $t$. The authors also compare schemes for certain functions with the corresponding classical ECCs and show that function correction can be done with a lower redundancy length than what is required for classical error correction. FCCs for symbol-pair read channels were explored in \cite{pairwise read}. Bounds on the optimal redundancy is provided and a counterpart of locally binary functions is proposed - pair-locally binary functions.  Further generalizations and improvements of the results in \cite{FCC} have been recently reported in \cite{SSY} \cite{GXZZ} \cite{LyS} \cite{SaR} \cite{RRHH} and \cite{VSS}.

In this paper, we propose some theoretical bounds on the redundancy of an FCC. Classical ECCs can be seen as a special case of FCCs obtained by considering the function to be a bijection. This gives us an approach to obtain bounds on the redundancy of systematic ECCs. A lower bound on the redundancy of systematic ECCs is also a bound for linear codes. Some of the well-known bounds for ECCs are the Hamming (Sphere Packing) bound, the Singleton bound, the Johnson bound, the Plotkin Bound and the Griesmer bound {\cite{coding theory}} . Out of these, the Griesmer bound is for linear codes and all the others are for non-linear codes. The theoretical bounds available for systematic/linear ECCs are much lower in number compared to the bounds available for non-linear codes. In {\cite{syst bound}}, Bellini, Guerrini and Sala (BGS) proposed a bound for non-linear codes as an improvement of a bound proposed by Zinoviev, Litsyn, and Laihonen (ZLL) {(\cite{ZLL 1},\cite{ZLL 2})}. The BGS bound and the ZLL bound were restricted to the case of systematic codes in {\cite{syst bound}}. {In this paper, we propose a bound on the redundancy of FCCs, by finding an upper bound on the independence number of the graph representation of FCCs proposed in {\cite{FCC}}. The bound is non-trivial in the region $2t+1 \le k$, where $k$ is the message length and $t$ is the error correction capability. By considering the function to be a bijection, we obtain a bound for systematic ECCs which is non-trivial in the region  $d\le k$, where $d$ is the minimum distance and $k$ is the message length.  We show that for single error correcting codes, our bound is at least as good as the ZLL bound. We compare it with the BGS bound for several parameters and show that our bound performs very close to the BGS bound.}

Furthermore, motivated by the importance of linear functions in modern engineering problems, we focus on FCCs where the function to be computed at the receiver is linear. In \cite{FCC}, the graph based representation is not made use of, as for a general function, the characterisation of this graph is difficult. For linear functions, we characterise this graph to get its spectrum, which leads to bounds on redundancy.  Furthermore, as explored in \cite{FCC}, the problem of function correction is more complex as compared to classical error correction due to the irregular nature of its distance requirement. Thus, a lower complexity coding scheme is proposed in \cite{FCC}, where the same parity is assigned to messages mapped to the same function value.  We identify a class of linear functions for which this type of coding is optimal, for certain parameters.

\subsection{Contributions and Organisation}
The technical contributions of this paper are summarised:
\begin{itemize}
	\item  {A function dependent graph based representation of the problem is proposed in {\cite{FCC}}. Independent sets of this graph, of sufficient cardinality can be chosen to be an FCC. By finding an upper bound for the independence number of this graph, we propose a lower bound on redundancy. We do this by considering a different graph with the same vertex set as the original one but with a maximum independent set which is a superset of that of the original graph. We express this graph as the Cartesian product of two smaller graphs which helps in characterising its independence number. This approach of bounding the independence number gives a non trivial bound in the region $2t+1\le k$. }(Section 
	\ref{sec: lb}: Theorem \ref{thm: Cartesian lower bound}).
	
	\item  By considering the function to be a bijection, we obtain a lower bound on redundancy for systematic ECCs. For the case when $2t+1\le k$, we identify  classes of codes that satisfy this bound with equality. We compare this bound with the ZLL bound and the BGS bound. We show that for single error correction, our bound and the ZLL bound are the same. We also compare the proposed bound with the best known bounds on linear codes for a range of parameters. (Section 
	\ref{subsec: systematic ECC}, \ref{subsec: comparison}).
	
	\item In \cite{FCC}, a Plotkin-like bound is derived for FCCs. The Hamming distance distribution of the message vector space is required for computing this bound. For linear functions, we show that only the weight distribution of the kernel of the function is needed to compute the bound (Section 
	\ref{sec: linear FCC}: Corollary \ref{thm:plotkin}).
	
	\item We characterise the graphical representation of FCCs for linear functions. We show that its adjacency matrix has a symmetric block circulant structure (Section \ref{sec: linear FCC}: Theorem \ref{linear adjacency}).
	
	\item Using this structure, we show that for linear functions, the
	adjacency matrix is diagonalised by an appropriate tensor powers of
	DFT matrices. As a special case, for linear functions whose
	domain is a vector space over a field of characteristic
	2, we show that Hadamard matrices can diagonalise the
	adjacency matrix. We use the spectrum of the graph thus obtained, to 
	get bounds on redundancy (Section \ref{sec: linear FCC}: Corollary \ref{cor:eigen val bound}).
	
	\item In \cite{FCC}, a construction of FCCs is proposed where the same parity vector is assigned to all message vectors that evaluate to the same
	function value. The redundancy obtained by using this method is an upperbound on the optimal redundancy. For linear functions, we term this construction coset-wise coding. We identify a class of linear functions for which the upper bound on redundancy obtained from coset-wise coding is tight and provide a coding scheme for such functions. (Section \ref{subsec: thm 5 tight}, \ref{subsec: thm 5 coding scheme}).

	\item We identify a class of linear functions for which the problem of coset-wise coding is equivalent to a lower dimensional classical error correction problem (Section \ref{subsec: eq to classical}).
\end{itemize}

\subsection{Notations}
The set of all positive integers less than or equal to $N$ is denoted by $[N]$. The set of all positive integers is denoted by $\mathbb{N}$ and the set of all non-negative integers is denoted by $\mathbb{N}_{0}$.  $\mathbb{C}$ stands for the set of complex numbers, and $\mathbb{R}$ for the set of real numbers. A finite field of size $q$ is represented as $\mathbb{F}_{q}$. The Hamming weight of $\mathbf{a} \in \mathbb{F}_{q}^{n}$ is denoted by $w_{H}(\mathbf{a})$. For $k\le l\le n$, $\mathbf{a}[k:l]$ represents the vector $(a_{k}a_{k+1}\hdots a_{l})$.
For any set of vectors $X \subseteq \mathbb{F}_{q}^{n}$, $w_{H}(X)$ denotes the Hamming weight of the minimum Hamming weight vector in $X$. For any pair of vectors $\mathbf{x},\mathbf{y} \in \mathbb{F}_{q}^{n}$, $d_H(\mathbf{x},\mathbf{y})$ denotes the Hamming distance between $\mathbf{x}$ and $\mathbf{y}$. For an $N\times N$ matrix $\mathbf{D}$, $[\mathbf{D}]_{ij}$ denotes the $(i,j)$th entry of $\mathbf{D}$ and $\mathbf{D}_{(i)}$ denotes the $i$th column of $\mathbf{D}$. For sets of positive integers ${X} ,{Y} \subseteq [N]$, $\mathbf{D}_{{X},{Y}}$ denotes the submatrix of $\mathbf{D}$ whose rows are indexed by ${X}$ and columns by ${Y}$. We use $\mathbf{e}_{i} \in \mathbb{F}_{q}^{n}$ to represent the unit vector with a $1$ in the $i$th coordinate and $0$s everywhere else. We use $\mathbf{I}_n$ to denote the identity matrix of order $n$. For a matrix $\mathbf{X}\in \mathbb{C}^{n\times n}$, $\mathbf{X}^\dagger$ denotes its conjugate transpose. For matrices $\mathbf{X}$ and $\mathbf{Y}$, $\mathbf{X}\otimes \mathbf{Y}$ denotes the tensor product of $\mathbf{X}$ and $\mathbf{Y}$. We use $\mathbf{X}^{\otimes n}$ to denote the $n$th order tensor power of $\mathbf{X}$ i.e., $\mathbf{X}^{\otimes n} = \otimes_{i=1}^n\mathbf{X}$. The set of all vectors that are at a Hamming distance less than or equal to $t$ from $\mathbf{u} \in \mathbb{F}_{q}
^{n}$ is denoted by $ B_H(\mathbf{u},t)$.
For any $x\in \mathbb{R}, [x]^+\coloneqq max(x,0)$. For any two integers $a$ and $n$,
\begin{equation*}
	<a>_n \coloneqq
	\begin{cases}
		
		a(mod\text{ }n);&\text{if } a(mod\text{ }n)\ne 0 \\
		n; &  \text{otherwise}.
	\end{cases} 
\end{equation*}

\section{Preliminaries}
\label{system}

In this section we briefly present the results, used subsequently, from graph theory, matrix theory and classical error correcting codes. Further, we present preliminaries on FCCs from \cite{FCC}, generalising  them to functions defined on vector spaces over $\mathbb{F}_q$

\subsection{Graphs}

A graph $\mathcal{G}$ is a pair $({V,E})$ where ${V}$ is the vertex set and ${E} \subseteq {V}\times {V}$ is the edge set. We say that vertices $v_1\text{ and }v_2$ are connected if $(v_1,v_2)\in E$. In this paper we consider simple graphs, i.e., there are no loops or multiple {edges. An independent set} of $\mathcal{G}$ is a set ${S}\subseteq {V}$ such that $v_{1},v_{2} \in {S} \implies (v_{1},v_{2}) \not\in {E}$.
The independence number $\alpha$ is the cardinality of the maximum independent set. A maximum independent set of $\mathcal{G}$ is called an $\alpha$-set of $\mathcal{G}$. The Cartesian product $\mathcal{G} \Osq \mathcal{H}$ of two graphs $\mathcal{G} = ({V},{E} )$ and $\mathcal{H} = ({V}',{E}')$ is a graph with its vertex set given by ${V} \times {V'}$ and two vertices $(u,v)$ and $(u',v')$ are connected in $\mathcal{G} \Osq \mathcal{H}$ iff 
\begin{enumerate}
	\item $u=u'$ and $(v,v')\in {E}'$ or,
	
	\item $(u,u') \in {E}$ and $v=v'$.
\end{enumerate}

\subsection{Circulant Matrices}
An $n\times n$ matrix $\mathbf{A}$ is said to be a circulant matrix if $[\mathbf{A}]_{ij} =[\mathbf{A}]_{<i+1>_n<j+1>_n} \text{ for all } i,j\in[n]$. An $nm\times nm$ matrix $\mathbf{B}$ is said to be a block circulant matrix if it is of the following block matrix form:

$$
\mathbf{B} = \begin{bmatrix}
	\mathbf{B}_{11}&\mathbf{B}_{12}&\mathbf{B}_{13}&\hdots&\mathbf{B}_{1n}\\
	\mathbf{B}_{21}&\mathbf{B}_{22}&\mathbf{B}_{23}&\hdots&\mathbf{B}_{2n}\\
	\vdots&\vdots&\vdots&\ddots&\vdots\\
	\mathbf{B}_{n1}&\mathbf{B}_{n2}&\mathbf{B}_{n3}&\hdots&\mathbf{B}_{nn}
\end{bmatrix},
$$where each $\mathbf{B}_{ij}$ is an $m\times m$ matrix and $\mathbf{B}_{ij}=\mathbf{B}_{_{<i+1>_n<j+1>_n}}$
for all $i,j\in[n]$. It is known that an $n\times n$ circulant matrix is diagonalized by the $n$th order Discrete Fourier Transform (DFT) matrix given by 
$$
\mathbf{W}_n=\small\begin{bmatrix}
	1&1&1&1&\hdots&1\\
	1&\omega&\omega^2&\omega^3&\hdots&\omega^{n-1}\\
	1&\omega^2&\omega^4&\omega^8&\hdots&\omega^{2(n-1)}\\
	\vdots&\vdots&\vdots&\ddots&\vdots\\
	1&\omega^{n-1}&\omega^{2(n-1)}&\omega^{3(n-1)}&\hdots&\omega^{(n-1)(n-1)}\\
\end{bmatrix},
$$
where $\omega = e^{i\frac{2\pi}{n}}$.
\normalsize

\subsection{Classical Error Correcting Codes}

An $(n,M,d)_q$ code $\mathcal{C}$ is a set of $M$ vectors of length $n$ with elements from some finite set (alphabet) of cardinality $q$ (say $\mathbb{F}_q$), such that $d$ is the minimum Hamming distance between any pair of vectors (codewords) in $\mathcal{C}$. A code of alphabet size $q$ is said to be a $q$-ary code. A linear code over $\mathbb{F}_q$ is a subspace of  $\mathbb{F}_q^n$. A linear code over $\mathbb{F}_q$ of dimension $k$ and distance $d$ is said to be an $[n,k,d]_q$ code.
A code of minimum distance $d$ can correct any error of length $\floor{\frac{d-1}{2}}$. The maximum number of codewords in any $q$-ary code of length $n$ and minimum distance $d$ between codewords is denoted by $A_q(n,d)$.	
The Hamming bound is given below:
\begin{thm}
	\cite{coding theory} For a $q$-ary code of length $n$ and minimum distance $d$, the maximum size of the code is upper bounded as,
	$$
	A_q(n,d) \le \frac{q^n}{\sum_{i=0}^{\floor{\frac{d-1}{2}}}\binom{n}{i}(q-1)^i}.
	$$
\end{thm}

Codes that satisfy the Hamming bound are called perfect codes. Hamming codes are linear single error correcting (minimum distance 3) perfect codes. The paramteres of a $q$-ary Hamming code $|\mathcal{C}|$ are $n = \frac{q^m-1}{q-1}$ and $k = \frac{q^m-1}{q-1}-m$ where $m$ is a positive integer such that $n$ and $k$ are positive.

The Singleton bound is given below:
\begin{thm}
	\cite{coding theory} For a $q$-ary code of length $n$ and minimum distance $d$, the maximum size of the code satisfies,		
	$$
	A_q(n,d) \le q^{n-d+1}.
	$$
\end{thm} 
Linear codes that satisfy the Singleton bound are called Maximum Distance Separable (MDS) codes. A $q$-ary MDS code of dimesnion $k$ and distance $d$ will be a $[k+d-1,k,d]_q$ code. It is believed that in the range $2\le k\le q-1$, MDS codes exist if and only if $k+d-1 \le q+1$ \cite{Roth}. For $k=1$ and $k=n$, trivial MDS codes exist.
\subsection{Linear Functions}

\label{defn:linear function}
A function $f : \mathbb{F}_{q}^{k} \to \mathbb{F}_{q}^{l} $ is said to be linear if it satisfies the following condition: $f(\alpha \mathbf{x} + \beta \mathbf{y}) = \alpha f(\mathbf{x}) +\beta f(\mathbf{y}), \forall\hspace{0.1cm} \mathbf{x},\mathbf{y} \in \mathbb{F}_{q}^{k} 
\hspace{0.15cm}\text{ and }\alpha , \beta \in  \mathbb{F}_{q}$. It can be expressed as a matrix operation $f(\mathbf{x}) = \mathbf{F}\mathbf{x}, \text{ for some }\mathbf{F} \in \mathbb{F}_{q}^{l \times k}.$

The kernel of $f$ or the null space of $\mathbf{F}$ is denoted by $ker(f)$. For the rest of the paper, the considered function is such that $l \le k$ and $\mathbf{F}$ is full rank, i.e., $rank_{\mathbb{F}_{q}}(\mathbf{F})=l$.
We denote the range of $f$ as $Im(f) = \mathbb{F}_{q}^{l} \triangleq \{f_{0},f_{1},\hdots f_{q^{l}-1}\}$.

\begin{defn}
	For a linear function  $f:\mathbb{F}_{q}^{k} \to \mathbb{F}_{q}^{l}$, we define $|ker(f)|_d$ to be the number of vectors in $ker(f)$ of Hamming weight $d$ and $\mathbb{F}_{q}^{k}/ker(f)$ to be the set of all cosets of $ker(f)$ in $\mathbb{F}_{q}^{k}$.
\end{defn}

\begin{rem}
	\label{rem:cosets}
	For a linear function $f : \mathbb{F}_{q}^{k} \to \mathbb{F}_{q}^{l}$, the cosets in $\mathbb{F}_{q}^{k}/ker(f)$ partition $\mathbb{F}_{q}^{k}$, where the elements of each coset are assigned to a unique function value, i.e there is an induced isomorphism $\tilde{f} : \mathbb{F}_{q}^{k}/ker(f) \to \mathbb{F}_{q}^{l}$ such that $	\mathbf{u} + ker(f) \mapsto f(\mathbf{u})$.
	We use $\bar{f_{i}}$ to denote the coset mapped to $f_{i}\in \mathbb{F}_{q}^{l}$ and $\bar{f_{0}}$ to denote $ker(f)$. Due to linearity, the distance distribution of $\bar{f_{i}} \text{ } \forall i \in \mathbb{F}_{q}^{l}$ is the same and is equal to the weight distribution of $ker(f)$. For non-linear functions, we use  $\bar{f_{i}}$ to denote the set of vectors in $\mathbb{F}_{q}^{k}$ that are mapped to $f_{i}\in Im(f)$.
\end{rem}

\subsection{Function-Correcting Codes}
\begin{figure}[H]
	\centering
	\includegraphics[scale=0.45]{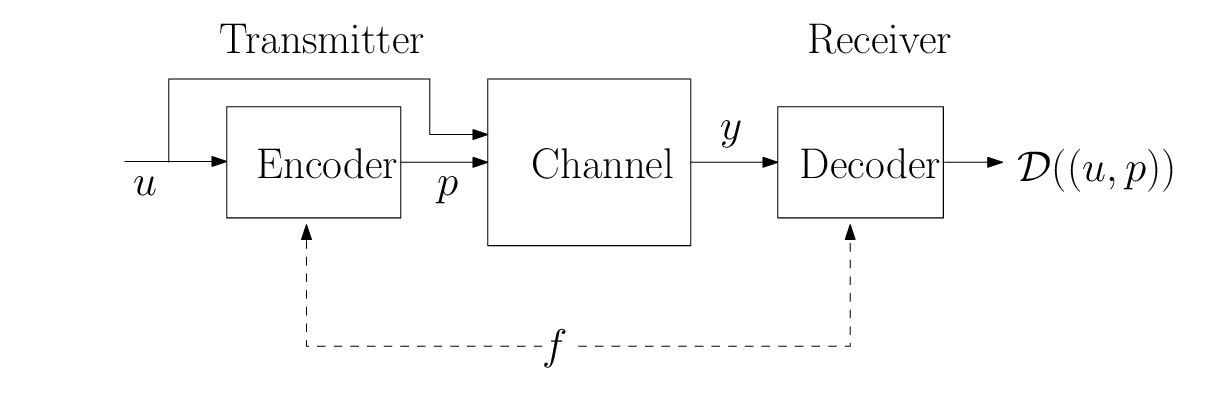}
	\caption{The function Correction Setting}
	\label{figure1}
\end{figure}

The system model illustrated in Fig. \ref{figure1} consists of a transmitter that wants to send a vector $\mathbf{u} \in \mathbb{F}_{q}^{k}$. The receiver wants to evaluate a function $f : \mathbb{F}_{q}^{k} \to Im(f) $ at $\mathbf{u}$. 
The transmitter encodes the data using a systematic encoding
$ \mathfrak{E} : \mathbb{F}_{q}^{k} \to \mathbb{F}_{q}^{k+r}$ such that $\mathfrak{E}(\mathbf{u}) = (\mathbf{u},\mathbf{p})$. The encoded data $\mathfrak{E}(\mathbf{u})$ is sent over a channel that can introduce {up to} $t$ symbol errors. The receiver receives $\mathbf{y} = \mathfrak{E}(\mathbf{u})+\mathbf{e} \in \mathbb{F}_{q}^{k+r} ,w_{H}(\mathbf{e}) \le t.$
{The receiver has a decoding function $\mathfrak{D} : \mathbb{F}_{q}^{k+r} \to Im(f),$ 
} which satisfies $\mathfrak{D}(\mathbf{y}) = f(\mathbf{u}) \hspace{0.15cm} \forall \hspace{0.15cm} u \in \mathbb{F}_{q}^{k}.$

The following definitions were given in \cite{FCC}, for FCCs over the binary field. We state them for any finite field $\mathbb{F}_q$. Moreover we state the results from \cite{FCC} taking $Im(f)$ to be $\mathbb{F}^l$ whenever linear functions are dealt with.
\begin{defn}
	\label{defn:fcc}
	\cite{FCC} A {systematic} encoding $\mathfrak{E} : \mathbb{F}_{q}^{k} \to \mathbb{F}_{q}^{k+r} $ is said to be an $(f, t)$-FCC for a function $f : \mathbb{F}_{q}^{k} \to Im(f)$ if for all $\mathbf{u}_{i},\mathbf{u}_{j} \in \mathbb{F}_{q}^{k}$ with $f(\mathbf{u}_{i}) \ne f( \mathbf{u}_{j})$, we have
	\begin{equation}
		d_{H}(\mathfrak{E}(\mathbf{u}_{i}),\mathfrak{E}(\mathbf{u}_{j})) \ge 2t+1.
	\end{equation}  
\end{defn}

\begin{rem}
	When the function $f:\mathbb{F}_q^k \to Im(f)$ is a bijection, an $(f,t)$-FCC is equivalent to a systematic $(n,q^k,2t+1)$ code. 
\end{rem}
\begin{defn}
	\label{optimal}
	\cite{FCC}
	The optimal redundancy $r_{f}(k,t)$ is defined as the smallest $r$ possible such that there exists an $(f, t)$-FCC with an encoding function $\mathfrak{E} : \mathbb{F}_{q}^{k} \to \mathbb{F}_{q}^{k+r}$.
\end{defn}

The following distance requirement matrix (DRM) was introduced in \cite{FCC}.
\begin{defn}
	\label{Distance Matrix}
	\cite{FCC}
	Let, $\mathbf{u}_{0}, \mathbf{u}_{1} \hdots \mathbf{u}_{q^{k}-1} \in \mathbb{F}_{q}^{k}$.  
	The distance requirement matrix (DRM) $\mathbf{D}_{f}(t,\mathbf{u}_{0}, \mathbf{u}_{1} \hdots \mathbf{u}_{q^{k}-1})$ for an $(f, t)$-FCC is a $q^{k} \times q^{k}$ matrix with entries
	\begin{small}
		\begin{equation*}
			[\mathbf{D}_{f}(t,\mathbf{u}_{0}, \hdots \mathbf{u}_{q^{k}-1})]_{ij}=
			\begin{cases}
				
				[2t+1 - d_{H}(\mathbf{u}_{i},\mathbf{u}_{j})]^{+}; &\hspace{-0.4cm} \text{~if }f(\mathbf{u}_{i}) \ne f( \mathbf{u}_{j})\\
				0; & \hspace{-0.3cm}\text{otherwise}.
			\end{cases} 
		\end{equation*}
	\end{small}
\end{defn} 
Notice that the DRM depends on the ordering among the elements of $\mathbb{F}_q^k.$

\begin{rem}
	\label{rem:nz}
	For a linear function $f : \mathbb{F}_{q}^{k} \to \mathbb{F}_{q}^{l},\text{ the } ker(f)$ has a dimension of $k-l$ over $\mathbb{F}_{q}$. So, each column of $\mathbf{D}_{f}(t,\mathbf{u}_{0}, \hdots \mathbf{u}_{q^{k}-1})$ will have at least $q^{k-l}$  number of 0s, since the cosets in $\mathbb{F}_{q}^{k}/ker(f)$ are of the same cardinality and all the elements of a particular coset are mapped to the same element in $\mathbb{F}_{q}^{l}$.
\end{rem}

\begin{rem}
	\label{rem:columns}
	Since $\mathbb{F}_{q}^{k}$ is a vector space over $\mathbb{F}_{q}$, the distance distribution from any vector is same as the weight distribution. As a result, for any function $f: \mathbb{F}_q^k \to Im(f)$, the columns (and rows) of $\mathbf{D}_{f}(t,\mathbf{u}_{0}, \hdots \mathbf{u}_{q^{k}-1})$ are permutations of each other.
\end{rem}

Irregular-distance codes of constraint $\mathbf{D}$ were defined in \cite{FCC} as follows:
\begin{defn}\cite{FCC}
	Let $\mathbf{D}\in \mathbb{N}_{0}^{M\times M}$. Then ${P} = \{\mathbf{p}_{i} : i\in [M] \}$ is said to be an irregular-distance code of constraint $\mathbf{D}$ or a $\mathbf{D}$-code if there is an ordering of ${P}$ such that $d_{H}(\mathbf{p}_{i},\mathbf{p}_{j}) \ge [\mathbf{D}]_{ij} \forall i,j \in [M]$. Further, $N_{q}(\mathbf{D})$ is defined as the smallest integer $r$ such that there exists a $\mathbf{D}$-code of length $r$ over $\mathbb{F}_q$.
\end{defn}

{For the case where $\mathbf{D}= \mathbf{D}_{f}(t,\mathbf{u}_{0}, \mathbf{u}_{1} \hdots \mathbf{u}_{q^{k}-1})$, we have $P= \{\mathbf{p}_{i} : i=0,1,2,\cdots q^k-1\}$ and the encoding of an $(f,t)$-FCC can be constructed as  $\mathfrak{E}(\bf u_i) = (\bf u_i,p_i)$. }

It is shown in \cite{FCC} that the problem of constructing an FCC can be formulated as that of finding sufficiently large independent sets of the following graph. 
\begin{defn}
	\cite{FCC}
	For a function $f:\mathbb{F}_q^k \to Im(f)$, $\mathcal{G}_{f}(t,k,r)$ is defined as the graph with vertex set $V = \mathbb{F}_q^k \times \mathbb{F}_q^r$, such that any two vertices $\mathbf{v}_{i} = (\mathbf{u}_{i},\mathbf{r}_{i})$ and $\mathbf{v}_{j} = (\mathbf{u}_{j},\mathbf{r}_{j})$ are connected if and only if
	\begin{itemize}
		\item $\mathbf{u}_{i} = \mathbf{u}_{j}$, or
		\item $f(\mathbf{u}_{i}) \ne f(\mathbf{u}_{j})$ and {$d_{H}(\mathbf{v}_{i},\mathbf{v}_{j}) < 2t+1.$}
	\end{itemize}
	
\end{defn}
That is, if two vertices {$(\mathbf{u_i},\mathbf{r_i}) \text{ and } (\mathbf{u_j},\mathbf{r_j})$} are connected in $\mathcal{G}_f(t,k,r)$, then an $(f,t)$-FCC cannot contain both $(\mathbf{u_i},\mathbf{r_i}) \text{ and } (\mathbf{u_j},\mathbf{r_j})$ as codewords. So if we can find an independent set of the graph of size $q^k$, it can be used as an $(f,t)$-FCC.
\begin{defn}
	\label{adjacency matrix}
	Let the vertex set of $\mathcal{G}_{f}(t,k,r)$ be $V = \{\mathbf{v}_1,\mathbf{v}_2,\hdots,\mathbf{v}_{q^{k+r}}\}$.
	The adjacency matrix $\mathbf{G}$ of the graph 	$\mathcal{G}_{f}(t,k,r)$ is a $q^{k+r}\times q^{k+r}$ matrix such that
	\begin{equation*}
		[\mathbf{G}]_{i,j} =
		\begin{cases}
			
			1;&\text{if } \mathbf{v}_{i}\text{ and }\mathbf{v}_{j} \text{ are connected} \\
			0; &  \text{otherwise}.
		\end{cases} 
	\end{equation*}
\end{defn}
Let $\gamma_{f}(k, t)$ denote the smallest integer $r$ such that there
exists an independent set of size $q^{k}$
in $\mathcal{G}_{f}(t,k,r)$.
\begin{figure}
	\centering
	\includegraphics[scale=0.75]{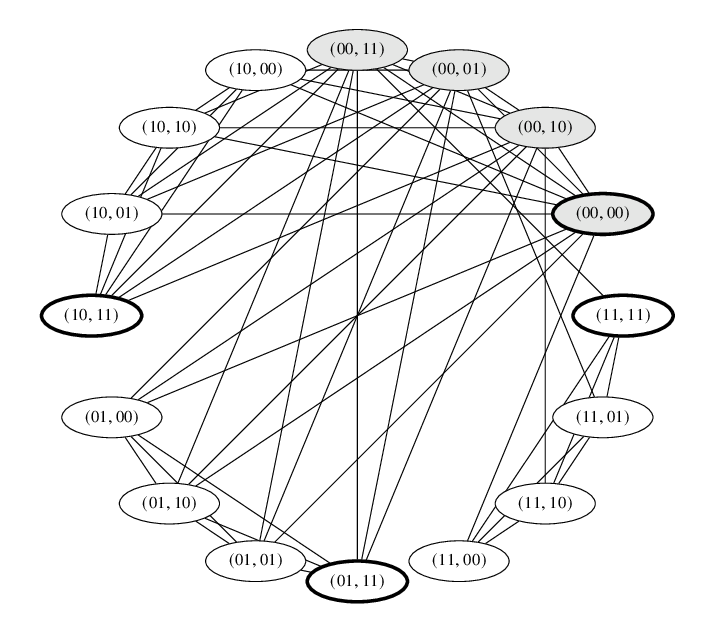}
	\caption{\cite{FCC} Graph $\mathcal{G}_f(t,k,r)$ for $t=1, k=2$ and $r=2$ and the function {$f((u_1u_2))=u_1\lor u_2$}. An independent set of size 4 is highlighted in bold.}
	\label{figure2}
\end{figure}

\begin{exmp}
	\label{exmp:fcc graph}
	Consider the function $f:\mathbb{F}_{2}^{2}\to \{0,1\}$ defined as {$f((u_{1}u_{2})) = u_{1}\lor u_{2}$}, where $\lor$ denotes the logical OR operator. The graph $\mathcal{G}_f(t,k,r)$ for $t$=1 and $r=2$ is shown in Fig. \ref{figure2}. This graph is a redrawn version of the graph provided in \cite{FCC}.
\end{exmp}

The following theorem from \cite{FCC} shows the connection between the redundancy of optimal FCCs, irregular-distance codes and independent sets of $\mathcal{G}_{f}(t,k,r)$.
\begin{thm}
	\label{thm:2t bound}
	\cite{FCC} For any function $f:\mathbb{F}_{q}^{k} \to Im(f)$,    
	\begin{equation}
		r_{f}(k,t) = \gamma_{f}(k, t) = N_q(\mathbf{D}_{f}(t,\mathbf{u}_{0}, \hdots \mathbf{u}_{q^{k}-1})).
	\end{equation}  
\end{thm}

{Thus, the problem of finding an optimal $(f,t)$-FCC is equivalent to finding an optimal irregualr distance code with DRM $\mathbf{D}= \mathbf{D}_{f}(t,\mathbf{u}_{0}, \mathbf{u}_{1} \hdots \mathbf{u}_{q^{k}-1})$, as well as finding the maximal independent set of the graph $\mathcal{G}_{f}(t,k,r)$.}

The definitions for function distance and function distance matrix (FDM) as in \cite{FCC} are given below:
\begin{defn}
	\cite{FCC}
	\label{defn:function distance }
	For a function $f:\mathbb{F}_{q}^{k} \to Im(f)$, the distance between $f_{i}, f_{j} \in Im(f)$ is defined as 
	\begin{equation}
		d_f(f_{i},f_{j})\triangleq\min_{\mathbf{u}_{i},\mathbf{u}_{j} \in \mathbb{F}_{q}^{k}}d_H(\mathbf{u}_{i},\mathbf{u}_{j}), \hspace{0.15cm} s.t.\hspace{0.15cm}f(\mathbf{u}_{i})=f_{i}, f(\mathbf{u}_{j})=f_{j}.
	\end{equation}
\end{defn}
\begin{defn}
	\cite{FCC}
	\label{defn:function distance  matrix (FDM)}
	The function distance matrix of a function $f:\mathbb{F}_{q}^{k} \to Im(f)$  is a $|Im(f)|\times |Im(f)|$ matrix with its entries given by
	$
		[\mathbf{D}_{f}(t,f_{0}, f_{1} \hdots f_{|Im(f)|-1})]_{ij} = $
		$$
		\begin{cases}
		
		[2t+1 - d_f(f_{i},f_{j})]^{+};&\text{if }i \ne j\\
			0; &  \text{otherwise}.
		\end{cases} 
	$$
\end{defn}

\begin{defn}	
	{Consider a linear function $f:\mathbb{F}_{q}^{k} \to \mathbb{F}_{q}^{l}$ with $\{\bar{f_{0}}, \bar{f_{1}},\hdots, \bar{f}_{q^l-1}\} \text{ being the cosets in } \mathbb{F}_q^k/ker(f)$. The inter-coset distance between any two cosets $\bar{f_{i}}\text{ and }\bar{f_{j}}$ is defined as the distance between $f_i$ and $f_j$.}
\end{defn}

\begin{rem}
	\label{rem:coset leaders}
	It follows from Remarks \ref{rem:cosets} and \ref{rem:nz}, that for {linear} functions, the problem of finding function distances becomes that of finding inter-coset distances. We also have,$$
	d_f(f_{i},0) =w_{H}(\bar{f_{i}}) =\min_{\mathbf{u}\in \bar{f_{i}}}w_{H}(\mathbf{u}).$$
	Because of the property of cosets, the inter-coset distance distributions are uniform, i.e. the column/row entries of 	$\mathbf{D}_{f}(t,f_{0}, f_{1} \hdots f_{q^{l}-1})$ are from the set $\{d_f(f_{i},0); f_{i} \in \mathbb{F}_{q}^{l} \}$, i.e, the columns (and rows) are permutations of each other.
\end{rem}
\begin{rem}
	\label{rem:equivalent functions}
	Consider linear functions $f$ and $g$ such that $f(\mathbf{x}) = \mathbf{F}\mathbf{x}$ and $g(\mathbf{x}) = \mathbf{G}\mathbf{x}$. Then the DRM (and FDM)  for $f$ and $g$ are the same if $\mathbf{F}$ can be obtained from $\mathbf{G}$ using only elementary row operations and column permutations.
\end{rem}
The following lower bound on redundancy is given in \cite{FCC}.
\begin{thm}
	\label{thm : df LB}
	\cite{FCC} For any function $f:\mathbb{F}_{q}^{k}\to Im(f)$ and $\{\mathbf{u}_{1}, \mathbf{u}_{2} \hdots \mathbf{u}_{M}\} \subseteq \mathbb{F}_{q}^{k}$
	\begin{equation}
		r_{f}(k,t) \ge N_{q}(\mathbf{D}_{f}(t,\mathbf{u}_{1}, \mathbf{u}_{2} \hdots \mathbf{u}_{M})
	\end{equation}  
	and for $|Im(f)| \ge 2$
	\begin{equation}
		r_{f}(k,t) \ge 2t.
	\end{equation}  
\end{thm}

As shown in \cite{FCC}, an existential bound on redundancy is obtained  when the same parity vector is assigned to all the messages that are mapped to the same function value.
\begin{thm}
	\label{thm : df UB}
	\cite{FCC} For any function $f:\mathbb{F}_{q}^{k} \to Im(f)$,  
	\begin{equation}
		r_{f}(k,t) \le N_q(\mathbf{D}_{f}(t,f_{0}, f_{1} \hdots f_{|Im(f)|-1}))
	\end{equation}  
	where $\mathbf{D}_{f}(t,f_{0}, f_{1} \hdots f_{|Im(f)|-1})$ is the FDM.
\end{thm}

The bound in Theorem \ref{thm : df UB} can be tight as shown in the following corollary which follows from Theorems \ref{thm : df LB} and \ref{thm : df UB}.
\begin{corollary}
	\label{label:optimal coding}
	If there exists a set of representative information vectors $\{\mathbf{u}_{0}, \mathbf{u}_{1} \hdots \mathbf{u}_{|Im(f)|-1}\}$ with $\{f(\mathbf{u}_{0})\hdots f(\mathbf{u}_{|Im(f)|-1})\} = Im(f)$ and 
	\begin{equation}
		\label{rep vectors}
		\mathbf{D}_{f}(t,f_{0}, f_{1} \hdots f_{|Im(f)|-1}) = \mathbf{D}_{f}(t,\mathbf{u}_{0}, \mathbf{u}_{1} \hdots \mathbf{u}_{|Im(f)|-1}),
	\end{equation}
	then
	\begin{equation*}
		r_{f}(k,t) = N_q(\mathbf{D}_{f}(t,f_{0}, f_{1} \hdots f_{|Im(f)|-1})).
	\end{equation*}

\end{corollary}
We propose a class of linear functions for which such a selection of representative  exists in Section \ref{subsec: thm 5 tight} 

The following generalized version of the Plotkin bound for binary FCCs is provided in \cite{FCC}.
\begin{lem}
	\label{plotkin}
	\cite{FCC} For any DRM $\mathbf{D} \in \mathbb{N}_{0}^{M\times M}$,
	\begin{equation*}
		N_{2}(\mathbf{D}) \ge 
		\begin{cases*}
			\frac{4}{M^{2}}\sum_{i,j:i<j}^{}[\mathbf{D}]_{ij}, &\text{if M is even}\\
			\frac{4}{M^{2}-1}\sum_{i,j:i<j}^{}[\mathbf{D}]_{ij}, &\text{if M is odd}.
		\end{cases*}
	\end{equation*}
\end{lem}

\begin{exmp}
	\label{exmp:explanation}
	Consider a function $f : \mathbb{F}_{2}^{4} \to \mathbb{F}_{2}^{2}$ with the mapping $$
	\mathbf{x} \mapsto \begin{bmatrix}
		1&1&1&0\\
		0&1&1&0
	\end{bmatrix}\mathbf{x}
	$$.

	The coset decomposition $\mathbb{F}_{2}^{4}/ker(f)$ and the induced mapping is: 
	\begin{align*}
		&\{0000, 0001, 0110, 0111\}\mapsto 00\text{ } (f_0)\\
		&\{0010, 0011, 0100, 0101\}\mapsto 11\text{ } (f_1)\\
		&\{1000, 1001, 1110, 1111\}\mapsto 10\text{ } (f_2)\\
		&\{1100, 1101, 1010, 1011\}\mapsto 01 \text{ }(f_3).
	\end{align*}
	{Let $\mathbf{u}_i$ be the $4$-bit binary representation of $i \in \{0,1,2 ,\hdots,15\}$.} For $t=2$, the DRM for an $(f,t)$-FCC is
	
	{$\mathbf{D}_{f}(t,\mathbf{u}_{0}, \mathbf{u}_{1} \hdots \mathbf{u}_{15}) =$}
	$$
	\footnotesize  \begin{bmatrix}
		0&0&4&3&4&3&0&0&4&3&3&2&3&2&2&1&\\
		0&0&3&4&3&4&0&0&3&4&2&3&2&3&1&2&\\
		4&3&0&0&0&0&4&3&3&2&4&3&2&1&3&2&\\
		3&4&0&0&0&0&3&4&2&3&3&4&1&2&2&3&\\
		4&3&0&0&0&0&4&3&3&2&2&1&4&3&3&2&\\
		3&4&0&0&0&0&3&4&2&3&1&2&3&4&2&3&\\
		0&0&4&3&4&3&0&0&2&1&3&2&3&2&4&3&\\
		0&0&3&4&3&4&0&0&1&2&2&3&2&3&3&4&\\
		4&3&3&2&3&2&2&1&0&0&4&3&4&3&0&0&\\
		3&4&2&3&2&3&1&2&0&0&3&4&3&4&0&0&\\
		3&2&4&3&2&1&3&2&4&3&0&0&0&0&4&3&\\
		2&3&3&4&1&2&2&3&3&4&0&0&0&0&3&4&\\
		3&2&2&1&4&3&3&2&4&3&0&0&0&0&4&3&\\
		2&3&1&2&3&4&2&3&3&4&0&0&0&0&3&4&\\
		2&1&3&2&3&2&4&3&0&0&4&3&4&3&0&0&\\
		1&2&2&3&2&3&3&4&0&0&3&4&3&4&0&0&\\
	\end{bmatrix}.
	$$
	Let $f_i \coloneq \mathbf{f}_i$ be the $2$-bit binary representation of $i \in \{0,1,2,3\}$. The FDM is given by

	$$
	{\mathbf{D}_{f}(t,f_{0}, f_{1}, f_{2}, f_{3})} =
	\begin{bmatrix}
		0&4&4&3\\
		4&0&3&4\\
		4&3&0&4\\
		3&4&4&0
	\end{bmatrix}.
	$$
	
	If we choose a set of representative vectors $\{\mathbf{u}_{0}, \mathbf{u}_{1},\mathbf{u}_{2},\mathbf{u}_{3}\} = \{0000,0100,1000,1100\}$ and $f_0 = 00,f_1 = 11,f_2 = 10,$ and $f_3 = 01,$ we can see that $\mathbf{D}_{f}(t,f_{0}, f_{1}, f_{2}, f_{3}) = \mathbf{D}_{f}(t,\mathbf{u}_{0}, \mathbf{u}_{1}, \mathbf{u}_{2}, \mathbf{u}_{3})$. So, by Corollary \ref{label:optimal coding}, $N_{2}(\mathbf{D}_{f}(t,f_{0}, f_{1}, f_{2}, f_{3}))=N_{2}(\mathbf{D}_{f}(t,\mathbf{u}_{0}, \mathbf{u}_{1} \hdots \mathbf{u}_{15}))$. That is, an optimal FCC can be obtained by finding parity vectors that satisfy the $4\times 4$ FDM $\mathbf{D}_{f}(t,f_{0}, f_{1}, f_{2}, f_{3})$, instead of the $16 \times 16$ DRM  $\mathbf{D}_{f}(t,\mathbf{u}_{0}, \mathbf{u}_{1} \hdots \mathbf{u}_{15})$. Thus the complexity of the problem is reduced.
	
\end{exmp} 

\section{A Lower Bound on Redundancy}
\subsection{Lower Bound on the Redundancy of FCCs }
\label{sec: lb}
In this section, we propose a lower bound on the redundancy of an FCC. We consider a graph $\mathcal\mathcal{G}'_{f}(t,k,r)$ with the same vertex set as of $\mathcal{G}_{f}(t,k,r)$ such that its maximum independent set contains the maximum independent set of $\mathcal{G}_{f}(t,k,r)$. This leads to an upper bound on $\alpha(\mathcal{G}_{f}(t,k,r))$, which gives a lower bound on the redundancy required. When the function considered is a bijection, we obtain a bound for classical systematic ECCs, which is examined in detail in the next subsection.

\begin{thm}
	\label{thm: Cartesian lower bound}
	{	For a function $f: \mathbb{F}_{q}^{k} \to Im(f)$ and $t \in \mathbb{N}$, there exists an $(f,t)$-FCC of redundancy $r$ only if}
	$$
	r \ge k - log({\alpha(\mathcal{G}_{f}(t,k,0))}).
	$$
\end{thm}

\begin{IEEEproof}
	Consider the graph $\mathcal{G}_{f}(t,k,0)$ with vertex set $\mathbb{F}_{q}^{k}$. Two vertices $\mathbf{x}_{1},\mathbf{x}_{2} \in \mathbb{F}_{q}^{k}$ are connected if and only if $f(\mathbf{x}_{1})\ne f(\mathbf{x}_{2})$ and $d_{H}(\mathbf{x}_{1},\mathbf{x}_{2})<2t+1$. Let $\mathcal{R}$ be the graph with vertex set $\mathbb{F}_{q}^{r}$ such that any two distinct vertices are connected. Note that $\alpha(\mathcal{R})=1$. Let $\mathcal{G}'_{f}(t,k,r)$ be the Cartesian product of the graphs $\mathcal{G}_{f}(t,k,0)$ and $\mathcal{R}$, i.e.,
	
	$$\mathcal{G}'_{f}(t,k,r) = \mathcal{G}_{f}(t,k,0) \Osq \mathcal{R}.$$An illustration is provided in Example \ref{exmp:cartesian product}.
	The vertex set of $\mathcal{G}_{f}(t,k,r)$ is the same as that of $\mathcal{G}'_{f}(t,k,r)$.
	
	Now, we show that if two vertices are unconnected in $\mathcal{G}_{f}(t,k,r)$, then they are also unconnected in $\mathcal{G}'_{f}(t,k,r)$. Let, $(\mathbf{x}_{i},\mathbf{v}_{i})$ and $(\mathbf{x}_{j},\mathbf{v}_{j}) \in \mathbb{F}_{q}^{k+r}$ be connected in $\mathcal{G}'_{f}(t,k,r)$. This happens if and only if \begin{enumerate}[label=(\roman*)]
		\item $\mathbf{x}_{i}=\mathbf{x}_{j}$, or 
		\item  ${f(\mathbf{x}_i)\ne f(\mathbf{x}_j),}$ $d_{H}(\mathbf{x}_{i},\mathbf{x}_{j})<2t+1$ and $\mathbf{v}_{i}=\mathbf{v}_{j}$.
	\end{enumerate} 
	Consider condition (i). If $\mathbf{x}_{i}=\mathbf{x}_{j}$, then $(\mathbf{x}_{i},\mathbf{v}_{i})$ and $(\mathbf{x}_{j},\mathbf{v}_{j})$ are connected in $\mathcal{G}_{f}(t,k,r)$ by definition. Consider condition (ii). If $d_{H}(\mathbf{x}_{i},\mathbf{x}_{j})<2t+1$ and $\mathbf{v}_{i}=\mathbf{v}_{j}$, then $d_{H}((\mathbf{x}_{i},\mathbf{v}_{i}),(\mathbf{x}_{j},\mathbf{v}_{i}))<2t+1$, which implies that $(\mathbf{x}_{i},\mathbf{v}_{i})$ and $(\mathbf{x}_{j},\mathbf{v}_{i})$ are connected in $\mathcal{G}_{f}(t,k,r)$. Hence, any $\alpha$-set of $\mathcal{G}_{f}(t,k,r)$ is a subset of some independent set of $\mathcal{G}'_{f}(t,k,r)$, which implies that 
	\begin{equation}
		\label{eqn:large indep set}
		\alpha(\mathcal{G}_{f}(t,k,r)) \le  \alpha(\mathcal{G}'_{f}(t,k,r)).
	\end{equation}
	
	Now, we bound the independence number of $\mathcal{G}'_{f}(t,k,r)$. Since $\mathcal{G}'_{f}(t,k,r) = \mathcal{G}_{f}(t,k,0) \Osq \mathcal{R}$, from \cite{vizing Cartesian} and \cite{product graphs}, we can use the upper bound on the independence number of a graph Cartesian product:

	\begin{align}
		\alpha(\mathcal{G}'_{f}(t,k,r) &\le min\{q^{k}\cdotp\alpha(\mathcal{R}), q^{r}\cdotp\alpha(\mathcal{G}_{f}(t,k,0))\}\nonumber\\
		\label{eqn:cart upper bound}
		&= min\{q^{k}, q^{r}\cdotp\alpha(\mathcal{G}_{f}(t,k,0))\}.
	\end{align}
	For an $(f,t)$-FCC of length $r$ to exist, $\alpha(\mathcal{G}_{f}(t,k,r))$ should be $q^{k}$. So, from (\ref{eqn:large indep set}) and (\ref{eqn:cart upper bound}), we can say that if an $(f,t)$-FCC of length $r$ exists, then $$q^{r}\cdotp\alpha(\mathcal{G}_{f}(t,k,0))\ge q^{k},$$
	which gives the required result.
\end{IEEEproof}
\begin{rem}
	{In the region $2t+1 > k$, $\alpha(\mathcal{G}_{f}(t,k,0)) = 1$, and the bound above simplifies to $r\ge k$. The RHS here is independent of $t$ and is a rather weak bound. This weakness comes from how the bound on the independence number of $\mathcal{G}_(t,k,r)$ is obtained - the connectivity of $(\mathbf{x}_{i},\mathbf{v}_{i})$ and $(\mathbf{x}_{j},\mathbf{v}_{i})$ in $\mathcal{G}'_(t,k,r)$ is dependent on $d_{H}(\mathbf{x}_{i},\mathbf{x}_{j})$ and not on  $d_{H}((\mathbf{x}_{i},\mathbf{v}_{i}),(\mathbf{x}_{j},\mathbf{v}_{j}))$. If $2t+1>k$, $\mathcal{G}_(t,k,0)$ becomes a fully connected graph and its independence number will be one, independent of $t$.}
\end{rem}

\begin{figure}\centering
	\subfloat[$\mathcal{R}=(\mathbb{F}_2, \{(0,1)\}$)]{\label{a}\includegraphics[width=.4\linewidth]{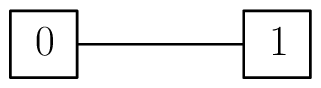}}\hfill
	\subfloat[$\mathcal{G}_f(t,k,0)$]{\label{b}\includegraphics[width=.6\linewidth]{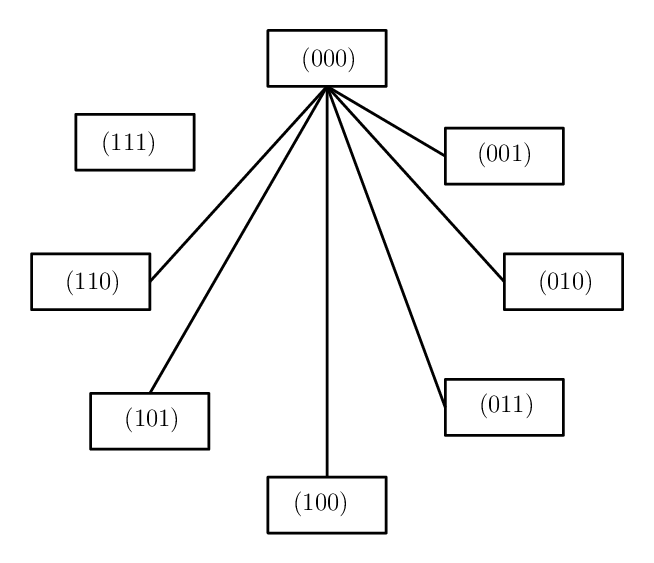}}\par 
	\subfloat[$\mathcal{G}'_{f}(t,k,r) = \mathcal{G}_{f}(t,k,0) \Osq \mathcal{R}$]{\label{c}\includegraphics[width=.7\linewidth]{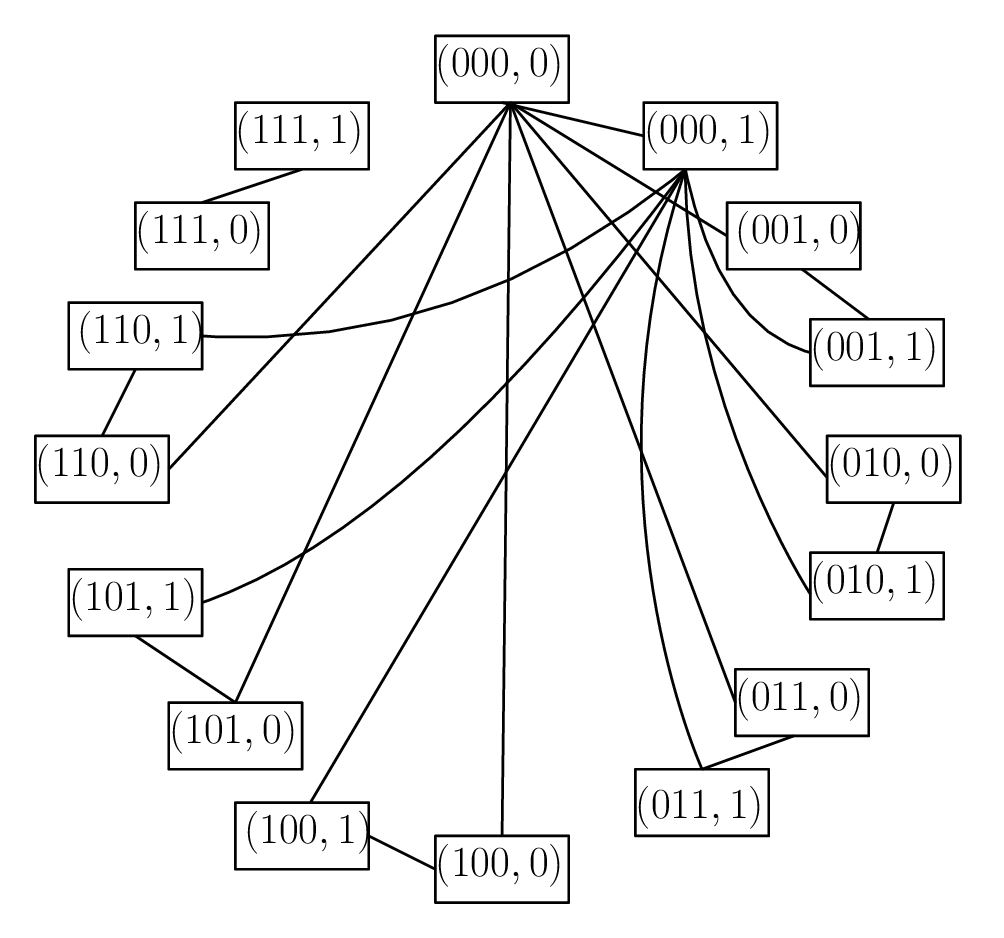}}
	\caption{Graphs $\mathcal{R}$, $\mathcal{G}_f(t,k,0)$ and $\mathcal{G}'_{f}(t,k,r)$ for $t=1$, $r=1$, $k=3$, and the function $f:\mathbb{F}_{2}^{3}\to \{0,1\}$ defined as {$f((u_{1}u_{2}u_3)) = u_{1}\lor u_{2}\lor u_3$}.}
	\label{figure3}
\end{figure}

\begin{exmp}
	\label{exmp:cartesian product}
	Consider the function $f:\mathbb{F}_{2}^{3}\to \{0,1\}$ defined as {$f((u_{1}u_{2}u_3)) = u_{1}\lor u_{2}\lor u_3$}. Let $t=1$. For $r=1$, consider the graph $\mathcal{R}$ with vertex set $\mathbb{F}_2$ with the two elements in $\mathbb{F}_2$ being connected. Consider the graph $\mathcal{G}_f(t,k,0)$ with vertex set $\mathbb{F}_2^3$ such that any two vertices $\mathbf{x}_i$ and $\mathbf{x}_j$ are connected if and only if $f(\mathbf{x}_i)\ne f(\mathbf{x}_j)$ and $d_H(\mathbf{x}_i,\mathbf{x}_j)<3$. Their Cartesian product $\mathcal{G}'_{f}(t,k,r) = \mathcal{G}_{f}(t,k,0) \Osq \mathcal{R}$ is shown in Fig. \ref{figure3}.
\end{exmp}
\begin{exmp}
	\label{exmp: 10-m}
	Consider the function $f:\mathbb{F}_2^{10} \to \mathbb{F}_2^m$ with the mapping $\mathbf{u} \mapsto \mathbf{Fu}$, where $\mathbf{F}=\left[\mathbf{I}_m\text{ } \mathbf{0}_{m\times (10-m)}\right]$, where $\mathbf{0}_{m\times (10-m)}$ is an $m \times (10-m)$ zero matrix. {For different values of $m$,  the lower bounds on the redundancy in bits for an $(f,t)$-FCC  are compared in Table {\ref{table:FCC comp}}. Since $f$ is linear and $\mathbf{F}$ has rank $m$, the partitions $f_0,f_1, \hdots f_{|Im(f)|} \subseteq \mathbb{F}_q^{k}$ each have a size of $2^{10-m}$.} We can see from Table \ref{table:FCC comp} that there are parameters for which the bound obtained from Theorem \ref{thm: Cartesian lower bound} is stronger. It can be seen that this bound is not always strong, as can been for the case $m=3, t=3.$ It is observed in general that the bound grows weaker as $2t+1$ approaches $k$.
	
	\begin{table*}
		
		\centering
		
		\begin{tabular}{|c|c|c|c|c|c|c|c|c|c|}
			\cline{2-10}
			\multicolumn{1}{c|}{}\multirow{2}{*}{}&\multicolumn{3}{c}{m=3}\vline&\multicolumn{3}{c}{m=6}\vline&\multicolumn{3}{c}{m=8}\vline\\
			\cline{2-10}
			\multicolumn{1}{c|}{}&$t=1$&$t=2$&$t=3$&$t=1$&$t=2$&$t=3$&$t=1$&$t=2$&$t=3$\\
			\hline
			Corollary 1\cite{FCC}&2&4&6&2&4&6&2&4&6\\
			\hline
			Lemma \ref{plotkin}\cite{FCC}&1&3&6&1&2&4&1&2&5\\
			\hline
			Theorem \ref{thm: Cartesian lower bound}&3&4&4&6&6&6&8&8&8\\
			\hline
		\end{tabular}
		\caption{Comparison of lower bounds on the redundancy for the $(f,t)$-FCC specified in Example \ref{exmp: 10-m}}
		\label{table:FCC comp}
	\end{table*}
\end{exmp}

The independence number of the graph $\mathcal{G}_{f}(t,k,0)$ is required to compute the bound in Theorem \ref{thm: Cartesian lower bound}. It is easier to compute $\alpha(\mathcal{G}_{f}(t,k,0))$ than $\alpha(\mathcal{G}_{f}(t,k,r))$ and the computation of the bound does not increase in complexity with $r$. From its definition, we can see that $\alpha(\mathcal{G}_{f}(t,k,0))$ is affected by the function $f$ and the error correction requirement $t$. As $t$ increases, the distance requirement for two vertices in the graph to be connected becomes weaker and thus $\alpha(\mathcal{G}_{f}(t,k,0))$ is a monotonically non-increasing function of $t$. Thus, the lower bound is a non-decreasing function of $t$. The distance distribution of the vectors in the partitions $f_0,f_1, \hdots f_{|Im(f)|} \subseteq \mathbb{F}_q^{k}$ and the sizes of these partitions affect $\alpha(\mathcal{G}_{f}(t,k,0))$. Example \ref{exmp: 10-m} shows how the lower bound is affected by $t$ and the partition sizes.

\subsection{Lower Bound on the Redundancy of Systematic ECCs}
\label{subsec: systematic ECC}
{In this subsection, we emphasise the usefulness of studying FCCs in order to study systematic ECCs - bounds on FCCs provide bounds on systematic ECCs, which is much scarcer in literature compared to the bounds available for general non-linear codes.}

A straightforward consequence of Theorem \ref{thm: Cartesian lower bound} is a lower bound on the redundancy required for classical systematic ECCs. This bound is obtained by considering the function $f$ to be a bijection.
\begin{corollary}
	\label{corr:classical lower bound}
	There exists a $({k+r},q^{k},d)_q$ systematic ECC only if
	\begin{equation}
		\label{ineq: Aq}
		{r} \ge k - log_q({A_{q}(k,d)}),
	\end{equation}
where $A_{q}(k,d)$ is the maximum number of codewords in any $q$-ary code of length $k$ and minimum distance $d$.
\end{corollary}

\begin{IEEEproof}
	Inequality (\ref{ineq: Aq}) is straightforward as $\alpha(\mathcal{G}_{f}(t,k,0)) = A_{q}(k,d)$ when the function $f$ is a bijection. 
\end{IEEEproof}
\begin{rem}
	{For the region $d > k$, $A_q(k,d) = 1$, and the bound above simplifies to a weak version - $r\ge k$.}
\end{rem}

Inequality (\ref{ineq: Aq}) shows that for a systematic $({k+r},q^{k},d)_q$ code with $d\le k$ to exist, the ratio of the size of the codomain of the encoding ($q^{k+r}$) to the size of the message space ($q^k$) should be at least the ratio of the size of the message space to the size of the largest code of distance $d$ in the message space. 

Corollary 2 means that obtaining an upperbound on $A_q(k,d)$ (a lower dimensional problem) will give  a lower bound on the redundancy of a $(k+r,q^k,d)$ systematic code.  An improvement in the upper bound of $A_q(k,d)$, improves the lower bound on redundancy given by Corollary \ref{corr:classical lower bound}. Moreover, the result shows that the framework of FCCs can be used to analyze systematic ECCs. Obtaining bounds on the independence number of $\mathcal{G}_f(k,t,0)$ will give us bounds on redundancy. 

Now, we discuss a range of parameters for which the bound is tight.

\begin{prop}
	Consider positive integers $n, q, 2\le k \le q-1, d$ with $d\le k$ such that there exists an $[n,k,d]_q$ MDS code. Then $r\coloneqq n-k, k$ and $d$ satisfy (\ref{ineq: Aq}) in Corollary \ref{corr:classical lower bound} with equality.
\end{prop}
\begin{IEEEproof}
	Consider a $[k+d-1,k,d]_q$ MDS code with $d\le k$. It is easy to see that then there exists  a $[k,k-d+1,d]_q$ MDS code. Thus, $A_q(k,d) = q^{k-d+1}$.
	So, inequality (\ref{ineq: Aq}) becomes $r\ge d-1$. The $[k+d-1,k,d]_q$ MDS code satisfies this with equality;
\end{IEEEproof}

\begin{rem}
	
	We now show that Hamming codes with $k\ge 3$ also achieve the minimum redundancy specified by Corollary \ref{corr:classical lower bound}. Consider an $[n,k,3]$ Hamming code over $\mathbb{F}_q$ where $n= \frac{q^r-1}{q-1}$ and $k=\frac{q^r-1}{q-1}-r \ge 3$. The bound from Corollary \ref{corr:classical lower bound} means that the redundacy $r= n-k$ is bounded as follows:
	\begin{equation}
		\label{ineqn parity}
		r \ge \ceil{k - log_q(A_q(k,3))}.
	\end{equation}
	For this bound to be tight, we have to show that $q^{r-1} < \frac{q^k}{A_q(k,3)}\le q^r$.  Since $A_q(k,3)$ can be bounded from above using the Hamming bound, we have 
	
	\begin{align*}
		\frac{q^k}{A_q(k,3)} &\ge \sum_{i=0}^{1}\binom{k}{i}(q-1)^i\\
		&= q^r - r(q-1).
	\end{align*}

	Now, we use an explicit linear code construction to get a lower bound on $A_q(k,3)$. Consider a parity check matrix $\mathbf{H}$ with $k$ columns and $r$ rows. For the code to have a minimum distance of $3$, the columns of $\mathbf{H}$ have to be non-zero and {distinct} columns have to be from distinct one dimensional subspaces of $\mathbb{F}_q^r$. The minimum number of rows in $\mathbf{H}$ required for this is $\ceil{log_q(k(q-1)+1)}$. Therefore, $A_q(k,3)\ge q^{k-\ceil{log_q(k(q-1)+1)}} = q^{k-\ceil{log_q(q^r - r(q-1))}}$. Thus, 
	$$
	\frac{q^k}{A_q(k,3)} \le q^{\ceil{log_q(q^r - r(q-1))}}.
	$$
	
	For all $q\ge 2, r\ge 2$ except $q=2$ and $r=2$, we can see that, 
	\begin{multline*}
		q^{r-1} < q^r - r(q-1) \le \frac{q^k}{A_q(k,3)}\le \\ q^{\ceil{log_q(q^r - r(q-1))}} \le q^r - r(q-1) \le q^r.
	\end{multline*}
	Thus, all $[n,k,3]_q$ Hamming codes with $k\ge 3$ satisfy (\ref{ineqn parity}) with equality. Bound (\ref{ineqn parity}) is not tight in general for perfect codes of minimum distance greater than 3. For $q=2, k=12$ and $d=7$, Corollary \ref{corr:classical lower bound} means that the redundancy $r\ge 6$, whereas the binary Golay code \cite{coding theory}, which is a perfect code  has a redundancy of 7.
	
\end{rem}

\subsection{Comparison with the ZLL Bound and the BGS Bound}
\label{subsec: comparison}

Now, we compare our bound with the ZLL bound and the BGS bound. The comparisons are for the case when $d\le k$. 

As given in \cite{syst bound}, the ZLL bound for systematic codes is as follows:
\begin{corollary}
	\label{corr:ZLL bound}
	(\cite{syst bound}, \cite{ZLL 1}, \cite{ZLL 2}) {Let $B_m^k$ denote the Hamming ball of radius $m$ in $\mathbb{F}_q^k$.} Let $k,d,m \in \mathbb{N}$, $d\ge 2, k\ge 1$. Let $n$ be such that there exists an $(n,q^k,d)_q$ systematic code. If $0\le m\le \floor{\frac{d-1}{2}}$, then
	\begin{equation}
		|B_m^k| \le A_q(n-k,d-2m).
	\end{equation}
	
\end{corollary}

The BGS bound for systematic codes, which is an improvement of the ZLL bound is as follows:
\begin{corollary}
	\label{corr:sys bound}
	\cite{syst bound} Let $k,d,m \in \mathbb{N}$, $d\ge 2, k\ge 1$. Let $n$ be such that there exists an $(n,q^k,d)$ systematic code. If $0\le m\le \floor{\frac{d-1}{2}}$, then
	
	\begin{equation}
		\label{eqn:BGS}
		|B_m^k| \le A_q(n-k,d-2m)-\frac{|B_m^{n-k}|}{|B_{d-2m-1}^{n-k}|} + 1,
	\end{equation}
	where $B_m^k$ denotes the Hamming ball of radius $m$ in $\mathbb{F}_q^k$.
\end{corollary}

For the case of single error correcting codes $(d=3)$, the ZLL bound simplifies to  
$$
r = {n-k} \ge log_q((q-1)k+1),
$$
which is the same as the inequality obtained by substituting the upper bound of $A_q(k,3)$ obtained from the Hamming bound in inequality (\ref{ineq: Aq}). This means that the bound in Corollary {\ref{corr:classical lower bound}} is at least as good as the ZLL bound for the $d=3 \le k$ case. 

\begin{figure}
	\centering
	\includegraphics{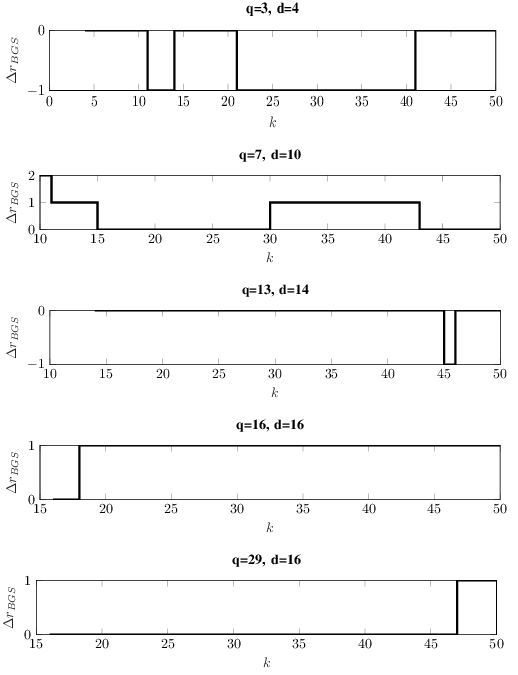}
	\caption{Plots of $\Delta r_{BGS} := r_{BGS}-r'$ vs $k$, for different values of field size $q$ and Hamming distance $d$, where $r_{BGS}$ is the lower bound on parity obtained using the BGS bound (Corollary \ref{corr:sys bound}) and $r'$ is the lower bound on parity obtained from Corollary \ref{corr:classical lower bound}.}
	\label{figure4}
\end{figure}

\begin{figure}
	\centering
	\includegraphics{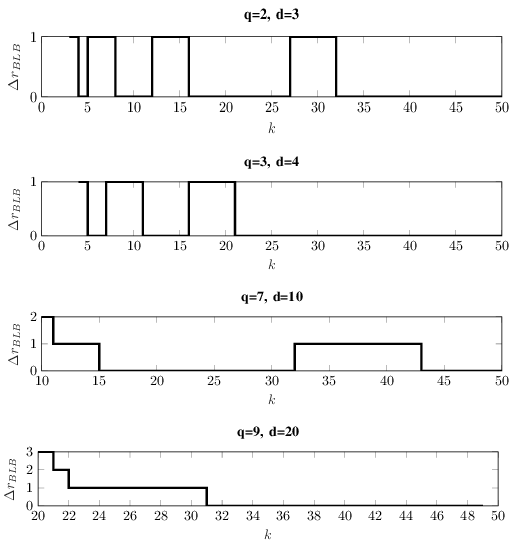}
	\caption{Plots of $\Delta r_{BLB} := r_{BLB}-r'$ vs $k$, for different values of field size $q$ and Hamming distance $d$, where $r_{BLB}$ is the best known lower bound on the parity of linear codes and $r'$ is the lower bound on parity obtained from Corollary \ref{corr:classical lower bound}.}
	\label{figure5}
\end{figure}

\begin{figure}
	\centering
	\includegraphics{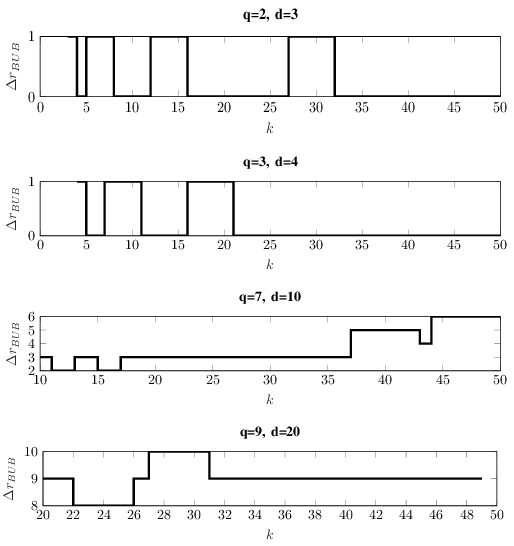}
	\caption{Plots of $\Delta r_{BUB} := r_{BUB}-r'$ vs $k$, for different values of field size $q$ and Hamming distance $d$, where $r_{BUB}$ is the best known upper bound bound on the parity of linear codes and $r'$ is the lower bound on parity obtained from Corollary \ref{corr:classical lower bound}.}
	
	\label{fig: comp upper}
\end{figure}

{An exact characterisation of the relationship between the proposed bound and the ZLL bound for $d\ne 3$, could not be obtained. Instead, we compare our bound with the BGS bound, an improved version of the ZLL bound, computationally for a wide range of parameters. It is an interesting open problem to characterise the parameters where one bound is stronger than the other. For that, we define $r_{BGS}$ as the lower bound on parity obtained using the BGS bound (Corollary {\ref{corr:sys bound}}) and $r'$ as the lower bound on parity obtained from Corollary {\ref{corr:classical lower bound}}. For a given $k$, we calculate $r_{BGS}$ as the smallest value of $n-k$ that satisfies ({\ref{eqn:BGS}}) for all $m \le \floor{\frac{d-1}{2}}$. We define $\Delta r_{BGS} \coloneq r_{BGS}-r'$, as the difference between the two bounds. Figure {\ref{figure4}} shows how $\Delta r_{BGS}$ varies with dimension $k$, for different values of Hamming distance $d$ and field size $q$. In the examples for field sizes $q = 3, 7, 13$ and $29$, we can see that $\Delta r_{BGS} = 0$, for a wide range of parameters, i.e., the two bounds perform the same in those regions. For $q = 3$ and $q=13$, we see that $\Delta r_{BGS} = -1$, i.e., the proposed bound performs better. But there are regions where the BGS bound outperforms the proposed bound. It is clear that there are areas where one bound is stronger than the other, even though we could not obtain an exact characterization.
	
	Since a lower bound on the parity of systematic codes is also a lower bound on the parity of linear codes, we also compare our lower bound (Corollary {\ref{corr:classical lower bound}}) with the best known lower bounds on the parity of linear codes, denoted as $r_{BLB}$. Figure {\ref{figure5}} shows how $\Delta r_{BLB} \coloneq r_{BLB}- r'$ varies with $k$ for different values of $d$ and $k$. We can observe that the proposed bound is as good as the best known lower bound for higher values of dimension $k$, i.e., $k >> d$. Values of $r_{BLB}$ were obtained from the MAGMA {\cite{MAGMA}} database. 
	
	Finally, we compare our bound with the best known upper bounds on the parity of linear codes, denoted as $r_{BUB}$. Figure {\ref{fig: comp upper}} shows how $\Delta r_{BUB} \coloneq r_{BUB}- r'$ varies with $k$ for different values of $d$ and $k$. Values of $r_{BUB}$ were also obtained from the MAGMA database. We observe that our bound is optimal for a large range of parameters, for lower field sizes. Since, either the values or strong bounds of $A_q(k,d)$ are known for a large range of parameters, the bound from Corollary {\ref{corr:classical lower bound}} is easy to compute.} 
{The reader should note that the proposed bound does not beat the previously known lower bounds. So, the best known lower bounds in the previous comparison in Figure {\ref{figure5}} are also optimal in the regions where the proposed bound is optimal.}

{Using this comparison, we emphasise the observation that Corollary {\ref{corr:classical lower bound}}, provides an easily computable lower bound on the redundancy of linear/systematic codes, that performs very close to the best known bounds, especially for larger dimensional cases. Thus, we see that the study of FCCs can lead to results about systematic ECCs. The weak nature of the bounds in Corollary {\ref{corr:classical lower bound}} and Theorem {\ref{thm: Cartesian lower bound}} for the region $2t+1 > k$, comes from how we tried to bound the independence number of $\mathcal{G}_f(t,k,r)$. Better estimates on the independence number will lead to better bounds on redundancy.}

\begin{rem}
	{In {\cite{FCC}}, the graphical representation for FCCs was proposed, but was not studied further to get bounds on redundancy. In {\cite{pairwise read}}, such a graphical approach was not used to define FCCs for symbol-pair read channels. The emphasis of this section is not only on the tightness of the bounds proposed for FCCs and systematic ECCs, but also on the usage of the graphical representation of FCCs proposed in {\cite{FCC}} to get bounds on FCCs and systematic ECCs.}
\end{rem}


\section{Linear-Function Correcting Codes  }
\label{sec: linear FCC}
In this section, we focus on codes for correcting linear functions. For linear functions, we show that the bound provided in Lemma \ref{plotkin} is simplified and can be obtained in terms of the weight distribution of the kernel of the function. Regarding the function-dependent graph $\mathcal{G}_f(t,k,r)$ we show that the adjacency matrix of it has a recursive block circulant structure for linear functions. For functions whose domain is a vector space over a finite field of characteristic $2$, we characterize the spectrum of the adjacency matrix, which leads to lower bounds on redundancy. A class of linear functions for which the upper bound on redundancy obtained from coset-wise coding is tight is identified (Section \ref{subsec: thm 5 tight}).

As mentioned in Remarks \ref{rem:nz} and \ref{rem:coset leaders}, linearity imposes a structure on the DRM and FDM of an FCC. This simplifies the bounds proposed in \cite{FCC} and gives rise to new bounds as shown in this section.

\subsection{Plotkin bound for FCCs for linear functions}
Now, for linear functions, we show that the Plotkin bound for FCCs provided in Lemma \ref{plotkin} can be simplified and can be made more computationally efficient to calculate. 
\begin{corollary}
	\label{thm:plotkin}
	For a linear function $f:\mathbb{F}_{q}^{k} \to \mathbb{F}_{q}^{l}$, the optimal redundancy of an $(f, t)$ FCC
	\begin{equation*}
		r_{f}(k, t) \ge \left(\frac{q}{q-1}\right)(2t+1)(1-q^{-l})-k+\frac{s}{(q-1)q^{k-1}},
	\end{equation*}  
	where $s = \sum_{x\in ker(f)}w_{H}(x) $ i.e, the sum of Hamming weights of the vectors in $ker(f)$.
\end{corollary}
\begin{IEEEproof}
	The bound can be obtained by counting $ \sum_{i\le j}^{}d_H(\mathbf{p}_{i},\mathbf{p}_{j})$ in two different ways.
	Let $\mathbf{D}$ =  $\mathbf{D}_{f}(t,\mathbf{u}_{0}, \hdots \mathbf{u}_{q^{k}-1})$ be the DRM for the function $f$.
	
	Let $\mathbf{p}_{1}, \mathbf{p}_{2}, \hdots \mathbf{p}_{q^{k}}$ be the codewords of a $\mathbf{D}$-code of length r. Arrange the codewords as a $q^{k} \times r$ matrix $\mathbf{P}$. Since each column of $\mathbf{P}$ can contribute at most $q^{2k}(1 - \frac{1}{q})$ to the sum $\sum_{i, j}^{}d_{H}(\mathbf{p}_{i},\mathbf{p}_{j})$, we have 
	\begin{equation}
		\label{plotkin 1}
		\sum_{i, j}^{}d_{H}(\mathbf{p}_{i},\mathbf{p}_{j}) \le rq^{2k-1}(q-1).
	\end{equation}
Furthermore, by definition,
	\begin{equation*}
		d_{H}(\mathbf{p}_{i},\mathbf{p}_{j}) \ge [\mathbf{D}]_{ij}, 
	\end{equation*}
	which implies that
	\begin{equation}
		\label{plotkin 2}
		\sum_{i, j}^{}d_{H}(\mathbf{p}_{i},\mathbf{p}_{j}) \ge \sum_{i, j}^{}[\mathbf{D}]_{ij}.
	\end{equation}
	Note that dim$_{\mathbb{F}_{q}}(ker(f)) = k-l$. Thus, there will be at least $q^{k-l}$ number of $0$s in each column. Because of linearity, the sum of non-zero entries of each column will be the same. Thus,
	\begin{align}
		\sum_{i, j}^{}[\mathbf{D}]_{ij} &= {\text{(no. of columns)} \times \text{(sum of one column)}}. \nonumber
	\end{align}
	To find the sum of one column, consider the first column of $\mathbf{D}$. Let $I$ be the row indices of the non-zero entries in the first column of $\mathbf{D}$. Let the first column correspond to the all zero vector $\mathbf{0}$. From the definition of $\mathbf{D}_{f}(t,\mathbf{u}_{0}, \hdots \mathbf{u}_{q^{k}-1})$ and Remark \ref{rem:coset leaders}, we have
	\begin{align*}
		[\mathbf{D}]_{i1} \ge 2t+1 - d_{H}(\mathbf{u}_{i},\mathbf{0})\\
		&= 2t+1 - w_{H}(\mathbf{u}_{i}).
	\end{align*}
	This implies that
	\begin{equation*}
		\sum_{i}^{} [\mathbf{D}]_{i1} \ge (2t+1)(q^{k}-q^{k-l}) - \sum_{i \in I}^{} w_{H}(u_{i}).
	\end{equation*}
	The sum of weights of all the vectors in $\mathbb{F}_{q}^{k}$ can be found to be $k(q-1)q^{k-1}$. Let the sum of weights of the vectors in $ker(f)$ be $s$, which gives
	
	\begin{align*}
		\sum_{i}^{} [\mathbf{D}]_{i1} & \ge (2t+1)(q^{k}-q^{k-l}) - k(q-1)q^{k-1} + s.
	\end{align*}
	Thus,
	\begin{align}
		{\sum_{i, j}^{}[\mathbf{D}]_{ij}} &\ge {q^{k} \times \sum_{i}^{} [\mathbf{D}]_{i1}} \nonumber\\
		&= {q^{k}((2t+1)(q^{k}-q^{k-l}) - k(q-1)q^{k-1} + s)}\label{plotkin 3}.
	\end{align}
	From (\ref{plotkin 1}), (\ref{plotkin 2}) and (\ref{plotkin 3}), we get the required result.
\end{IEEEproof}
The Plotkin bound provided in \cite{FCC} requires the sum of all $[\mathbf{D}]_{ij}$ to calculate the bound. For linear functions, we can see that we require only the sum of the Hamming weights of the vectors in $ker(f)$.

\begin{rem}
	{For systematic ECCs, by choosing the function $f$ to be a bijection, i.e., $s=0$ and $l=k$, the bound in Corollary {\ref{thm:plotkin}} reduces to}
	
	\begin{equation*}
	{
			n := k + r_f(k,t)\ge \frac{(2t+1)(q^k-1)}{q^{k-1}(q-1)},
		}
	\end{equation*}
	{which is the same as the Plotkin bound for classical error correcting codes.}
\end{rem}

\subsection{Structure of $\mathcal{G}_{f}(t,k,r)$ for linear functions}

We now consider the structure linearity imposes on the graph $\mathcal{G}_{f}(t,k,r)$. We show that its adjacency matrix $\mathbf{G}$ has a recursive block circulant structure. For cases when the finite field size $q$ is prime, the rows and columns of $\mathbf{G}$ can be indexed by the vectors in $\mathbb{F}_q^{k+r}$ in lexicographic order. When $q$ is not a prime, such a lexicographic ordering is not possible. In that case we show that there is an ordering of the rows and columns of $\mathbf{G}$ in this recursive block circulant form. We exploit this recursive block circulant structure to get the spectrum of the adjacency matrix, which can be made use of to get lower bounds on the redundancy of the FCC.
\begin{defn}
	For  $\mathbf{u} \in \mathbb{F}_{q}^{n}$ and $m \in [n]\cup \{0\}$, define ${P}_{m}^n(\mathbf{u}) \triangleq \{(x_{1}x_{2}\hdots x_{n})\in \mathbb{F}_{q}^{n}: (x_{1}x_{2}\hdots x_{m})=(u_{1}u_{2}\hdots u_{m})\}$ if $m \ne 0$ and $P_{m}^n(\mathbf{u})= \mathbb{F}_{q}^{n}$ if $m=0$. 
	
\end{defn}
That is, $P_{m}^n(\mathbf{u})$ is the set of all $n$ length vectors over $\mathbb{F}_q$ such that its first $m$ indices is equal to the first $m$ indices of $\mathbf{u}$.
\begin{thm}
	\label{linear adjacency}
	For a linear function $f:\mathbb{F}_{q}^{k}\to\mathbb{F}_{q}^{l}$, let $\mathbf{G} \in \{0,1\}^{q^{n}\times q^{n}}$, where $n\coloneqq k+r$, be the adjacency matrix of the graph $\mathcal{G}_{f}(t,k,r)$. Let the rows and columns of the matrix $\mathbf{G}$ be indexed by the vectors in $\mathbb{F}_{q}^{n}$. Then, $\mathbf{G}$ satisfies the following condition:
	
	\begin{itemize}
		\item[$C1$] For all $\mathbf{u},\mathbf{v} \in \mathbb{F}_{q}^{n}$ and for all $m \in [n-1]\cup \{0\}$, for some ordering of  $P_{m}^n(\mathbf{u})\text{ and }P_{m}^n(\mathbf{v})$, the submatrix $\mathbf{S}\coloneqq \mathbf{G}_{P_{m}^n(\mathbf{u}),P_{m}^n(\mathbf{v})} \in  \{0,1\}^{q^{(n-m)}\times q^{(n-m)}}$,  expressed in block matrix form as
		\begin{equation*}
			\mathbf{S} = \begin{bmatrix}
				\mathbf{S}_{1,1} & \mathbf{S}_{1,2}& \hdots & \mathbf{S}_{1,q}\\
				
				\mathbf{S}_{2,1} & \mathbf{S}_{2,2}& \hdots & \mathbf{S}_{2,q}\\
				
				\vdots & \vdots& \ddots & \vdots\\
				\mathbf{S}_{q,1} & \mathbf{S}_{q,2}& \hdots & \mathbf{S}_{q,q}\\
			\end{bmatrix},
		\end{equation*}
		is block circulant, where $\mathbf{S}_{i,j}$ is of order $q^{n-m-1}$. That is, for any $i,j \in [q]$, \begin{equation}
			\mathbf{S}_{i,j} = \mathbf{S}_{<i+1>_q,<j+1>_q}  \forall\text{ }i,j\in[q]. \tag{A}
			\label{eqn:C1}
		\end{equation}
	\end{itemize}
\end{thm}
\begin{IEEEproof}
	{Consider the case when $q$ is a prime number $p$. We have to show that for all $m \in [n-1]\cup\{0\}$ and for all $\mathbf{u},\mathbf{v}\in \mathbb{F}_{p}^{n}$, the submatrix $\mathbf{G}_{P_{m}^n(\mathbf{u}),P_{m}^n(\mathbf{v})}$ satisfies (\ref{eqn:C1}) in condition $C1$. Assume that $P_{m}^n(\mathbf{u})$ and $P_{m}^n(\mathbf{v})$ are in lexicographic ordering. The first $m$ positions of the row and column indices of $\mathbf{S} = \mathbf{G}_{P_{m}^n(\mathbf{u}),P_{m}^n(\mathbf{v})}$ are the same. Because of the lexicographic ordering,  (\ref{eqn:C1}) in condition $C1$ can be expressed as: for any $\mathbf{q} \in P_{m}^n(\mathbf{u})$ and $\mathbf{r} \in P_{m}^n(\mathbf{v})$, }
	
	$$ {[\mathbf{G}_{P_{m}^n(\mathbf{u}),P_{m}^n(\mathbf{v})}]_{\mathbf{q},\mathbf{r}}=[\mathbf{G}_{P_{m}^n(\mathbf{u}),P_{m}^n(\mathbf{v})}]_{\mathbf{g},\mathbf{h}}}$$
	
	{where $\mathbf{g} = \mathbf{q}+a\cdot\mathbf{e}_{m+1}$ and $\mathbf{h}= \mathbf{r}+a\cdot\mathbf{e}_{m+1)}$\text{ }$\forall$\text{ } $a \in \mathbb{F}_{p}$. } Example \ref{exmp: indexing} shows the case for $q=3$ and $n=2$.

	{To show that $\mathbf{S}$ satisfies (\ref{eqn:C1}), it is now enough to show that for all $m \in [n]\cup\{0\}$ and for all $\mathbf{u},\mathbf{v}\in \mathbb{F}_{p}^{n}$, the submatrix $\mathbf{G}_{P_{m}^n(\mathbf{u}),P_{m}^n(\mathbf{v})}$ satisfies the following condition:}
	$$
	{[\mathbf{G}_{P_{m}^n(\mathbf{u}),P_{m}^n(\mathbf{v})}]_{\mathbf{q},\mathbf{r}}=0 \iff [\mathbf{G}_{P_{m}^n(\mathbf{u}),P_{m}^n(\mathbf{v})}]_{\mathbf{g},\mathbf{h}}=0},
	$$
	{where $\mathbf{g} = \mathbf{q}+a\cdot\mathbf{e}_{m+1}$ and $\mathbf{h} = \mathbf{r}+a\cdot\mathbf{e}_{m+1}$  for any $a \in \mathbb{F}_{p}$.} 
	
	{	From the definition of $\mathbf{G}$, we know that $[\mathbf{G}]_{\mathbf{q},\mathbf{r}}=0$ if and only if }
	\begin{enumerate}
		\item {$f(\mathbf{q}[1:k]) \ne f(\mathbf{r}[1:k])$ and $d_{H}(\mathbf{q},\mathbf{r}) \ge 2t+1$, or}
		\item {$\mathbf{q}[1:k] \ne \mathbf{r}[1:k]$ and $f(\mathbf{q}[1:k]) = f(\mathbf{r}[1:k])$.}
	\end{enumerate}
	{	Consider $\mathbf{q}$ and $\mathbf{r}$ satisfying the first condition, then, $[\mathbf{G}_{P_{m}^n(\mathbf{u}),P_{m}^n(\mathbf{v})}]_{\mathbf{q},\mathbf{r}}=0 \iff [\mathbf{G}_{P_{m}^n(\mathbf{u}),P_{m}^n(\mathbf{v})}]_{\mathbf{g},\mathbf{h}}=0$, as $\mathbf{g}[1:k]\text{ and }\mathbf{h}[1:k]$ lie in different cosets of $\mathbb{F}_{p}^{k}/ker(f)$ and $d_{H}(\mathbf{g},\mathbf{h}) = d_{H}(\mathbf{q},\mathbf{r})$. Now, consider $\mathbf{q}$ and $\mathbf{r}$ satisfying the second condition, i.e., $\mathbf{q}[1:k] \ne \mathbf{r}[1:k]$ and $\mathbf{q}[1:k]-\mathbf{r}[1:k] \in ker(f)$. Then, from the property of cosets, $\mathbf{g}[1:k]-\mathbf{h}[1:k] \in ker(f)$. That is, for such $\mathbf{q}$ and $\mathbf{r}$, $[\mathbf{G}_{P_{m}^n(\mathbf{u}),P_{m}^n(\mathbf{v})}]_{\mathbf{q},\mathbf{r}}=0 \iff [\mathbf{G}_{P_{m}^n(\mathbf{u}),P_{m}^n(\mathbf{v})}]_{\mathbf{g},\mathbf{h}}=0$. Thus, $\mathbf{G}$ satisfies condition $C1$.
		
		When $q$ is a prime power, i.e., $q = p^m$, for some prime $p$, $\mathbb{F}_q^k$ is isomorphic to the vector space $\mathbb{F}_p^{mk}$. So the same arguments can be considered where the matrix $\mathbf{G}$ is indexed by the vectors in $\mathbb{F}_p^{mk}$.}
\end{IEEEproof}
\begin{exmp}
	\label{exmp: indexing}
	Consider a function $f: \mathbb{F}_{3}^{2} \to \mathbb{F}_{3}^{2}$, defined as $f(\mathbf{u}) = \begin{bmatrix}
		1&1
	\end{bmatrix}\mathbf{u}$. For $r=0$ and $t=1$, the adjacency matrix $\mathbf{G}$ of the graph $\mathcal{G}_f(t,k,r)$ is 
	$$\mathbf{G}=
	\left[\begin{array}{ccc|ccc|ccc}
		0 &1 &1 &1 &1 &0 &1 &0 &0\\
		1 &0 &1 &0 &0 &1 &0 &0 &0\\
		1 &1 &0 &0 &0 &1 &0 &1 &0\\
		\hline
		1 &0 &0 &0 &0 &0 &1 &1 &0\\
		1 &0 &0 &0 &0 &1 &0 &1 &1\\
		0 &1 &1 &0 &1 &0 &1 &0 &1\\
		\hline
		1 &0 &0 &1 &0 &1 &0 &1 &0\\
		0 &0 &1 &1 &1 &0 &1 &0 &1\\
		0 &0 &0 &0 &1 &1 &0 &1 &0\\
	\end{array}\right].
	$$
	Consider the vectors $\mathbf{u}=00$ and $\mathbf{v}=10$. For $m=0$, $P_0^2(\mathbf{u})=P_0^2(\mathbf{v})=\mathbb{F}_3^2.$ Thus, $\mathbf{S} = \mathbf{G}_{P_0^2(\mathbf{u}),P_0^2(\mathbf{v})} = \mathbf{G}$. For convenience, for any non-negative integer $i \le p$, we use $\mathbf{i}_p^n$ to denote the $n$ length $p$-ary representation of $i$ (the $p$-ary representation of $i$ preceded by a sufficient number of zeros to make it length $n$). Thus $\mathbf{G}$ can be indexed using the integers as follows $[\mathbf{G}]_{ij} = [\mathbf{G}]_{\mathbf{i}_3^2,\mathbf{j}_3^2}$, where $i,j \in \{0,1,\hdots, 8\}$. It can be verified easily that    $[\mathbf{G}]_{\mathbf{i}_3^2,\mathbf{j}_3^2} = [\mathbf{G}]_{\mathbf{g}_3^2,\mathbf{h}_3^2}$, where $\mathbf{g}_p^n = \mathbf{i}_3^2 + a\cdot\mathbf{e}_{1}$ and $\mathbf{h}_3^2 = \mathbf{j}_3^2 + a\cdot\mathbf{e}_{1}$ for all $a \in \mathbb{F}_3$. This is equivalent to saying that $[\mathbf{G}]_{ij} = [\mathbf{G}]_{gh}$, where $g = i + b\cdot 3^1 (mod\text{ }9) $ and $h =  i + b\cdot 3^1 (mod\text{ }9)$ for all $b\in\{0,1,2\}$, leading to the block-circulant structure of $\mathbf{G}$.
	
	For $m=1$, $P_1^2(\mathbf{u})= \{00,01,02\}$ and $P_1^2(\mathbf{v})=\{10,11,12\}$, thus $\mathbf{S} = \mathbf{G}_{P_1^2(\mathbf{u}),P_1^2(\mathbf{v})} =\begin{bmatrix}
		1&1&0\\
		0&0&1\\
		0&0&1
	\end{bmatrix}$. This is the block in the first row and second column of the above block matrix representation of $\mathbf{G}$. It can be easily verified that for any $\mathbf{q}\in P_1^2(\mathbf{u})$ and $\mathbf{r}\in P_1^2(\mathbf{v})$, $ [\mathbf{G}_{P_1^2(\mathbf{u}),P_1^2(\mathbf{v})}]_{\mathbf{q},\mathbf{r}} = [\mathbf{G}_{P_1^2(\mathbf{u}),P_1^2(\mathbf{v})}]_{\mathbf{g},\mathbf{h}}$, where $\mathbf{g} = \mathbf{q} + a\cdot\mathbf{e}_2$ and $\mathbf{h} = \mathbf{r} + a\cdot\mathbf{e}_2$ for all $a \in \mathbb{F}_3$. This is equivalent to saying that $[\mathbf{S}]_{ij} = [\mathbf{S}]_{gh}$, where $g = i + b\cdot 3^0 (mod \text{ }3)$ and $h = j + b\cdot 3^0 (mod \text{ }3)$ for all $b\in\{0,1,2\}$, i.e., $\mathbf{S}$ is circulant.

\end{exmp}
\begin{rem}
	\label{rem:rec circ}
	Equivalently the structure of $\mathbf{G}$ can be seen as follows: express $\mathbf{G}$ in block matrix form as $$\mathbf{G} = \begin{bmatrix}
		\mathbf{G}_{1,1} & \mathbf{G}_{1,2}& \hdots & \mathbf{G}_{1,q}\\
		
		\mathbf{G}_{2,1} & \mathbf{G}_{2,2}& \hdots & \mathbf{G}_{2,q}\\
		
		\vdots & \vdots& \ddots & \vdots\\
		\mathbf{G}_{q,1} & \mathbf{G}_{q,2}& \hdots & \mathbf{G}_{q,q}\\
	\end{bmatrix}.
	$$Let $\mathbf{G}_{i} \coloneqq \mathbf{G}_{1,i}$ Then $\mathbf{G}$ will be of the form $$\mathbf{G} = \begin{bmatrix}
		\mathbf{G}_{1} & \mathbf{G}_{2}&\mathbf{G}_3& \hdots &\mathbf{G}_{q-1}&\mathbf{G}_q\\
		
		\mathbf{G}_{q} & \mathbf{G}_{1}&\mathbf{G}_{2}& \hdots  &\mathbf{G}_{q-2}&\mathbf{G}_{q-1}\\
		
		\mathbf{G}_{q-1} & \mathbf{G}_{q}&\mathbf{G}_{1}& \hdots  &\mathbf{G}_{q-3}&\mathbf{G}_{q-2}\\
		\vdots & \vdots&\vdots& \ddots &\vdots& \vdots\\
		\mathbf{G}_{3} & \mathbf{G}_{4}&\mathbf{G}_{5}& \hdots  &\mathbf{G}_{1}&\mathbf{G}_{2}\\
		\mathbf{G}_2 & \mathbf{G}_{3}&\mathbf{G}_{4}& \hdots  &\mathbf{G}_{q}&\mathbf{G}_{1}\\
	\end{bmatrix}.
	$$Note that the matrix $\mathbf{G}$ is symmetric as it is an adjacency matrix. Thus, $\mathbf{G}$ is a symmetric block circulant matrix. The submatrices $\mathbf{G}_1, \mathbf{G}_2 \hdots, \mathbf{G}_{q}$ can further be divided into their submatrices recursively and the block circulant structure will hold.
\end{rem}

\begin{corollary}
	\label{corr: power of 2}
	Consider the adjacency matrix $\mathbf{G} \in \{0,1\}^{2^{nm}\times 2^{nm}}$ corresponding to a function $f: \mathbb{F}_{2^m}^{k}\to \mathbb{F}_{2^m}^{l}$. Let the rows and columns of the matrix $\mathbf{G}$ be indexed by the set of vectors $\mathbb{F}_{2}^{nm}$ in lexicographic order. Then $\mathbf{G}$ satisfies the following condition: \begin{itemize}
		\item[{C$2$}] For all $\mathbf{u}_{i},\mathbf{v}_{i} \in \mathbb{F}_{2}^{i}$ and for all $i \in [nm]\cup \{0\}$, if the submatrix $\mathbf{S}\coloneqq \mathbf{G}_{U_{i}^{nm},V_{i}^{nm}}$ is expressed in block matrix form as
		$
		\mathbf{S} = \begin{bmatrix}
			\mathbf{S}_{1,1} & \mathbf{S}_{1,2}\\
			\mathbf{S}_{2,1} & \mathbf{S}_{2,2}
		\end{bmatrix},
		$
		where $\mathbf{S}_{i,j}\in \{0,1\}^{2^{n-m-1}\times 2^{n-m-1}} \text{ }\forall \text{ }i,j \in [2]$, then $\mathbf{S}_{1,1}$ = $\mathbf{S}_{2,2}$ and $\mathbf{S}_{1,2}=\mathbf{S}_{2,1}$.
	\end{itemize}
	
\end{corollary}
{Note that condition C$2$ is condition C$1$ with $q=2$.}
\begin{rem}
	Equivalently the structure of $\mathbf{G}$ can be seen as follows: express $\mathbf{G}$ in block matrix form as $\mathbf{G} = \begin{bmatrix}
		\mathbf{G}_{1,1} & \mathbf{G}_{1,2}\\
		\mathbf{G}_{2,1} & \mathbf{G}_{2,2}
	\end{bmatrix}.
	$
	Then $\mathbf{G}_{1,1} = \mathbf{G}_{2,2} \text{ and } \mathbf{G}_{1,2}=\mathbf{G}_{2,1}$ and the submatrices $\mathbf{G}_{1,1},   \mathbf{G}_{1,2}, \mathbf{G}_{2,1}\text{ and } \mathbf{G}_{2,2}$ can further be divided into their submatrices recursively and the above condition will hold. The upshot is that $\mathbf{G}$ can be fully specified using just its first row. The following example illustrates this 
\end{rem}

\begin{exmp}
	Consider a function $f: \mathbb{F}_{2}^{3} \to \mathbb{F}_{2}^{2}$, defined as $f(\mathbf{u}) = \begin{bmatrix}
		0&1&1\\
		1&1&0
	\end{bmatrix}\mathbf{u}$.
	For $t=1$ and $r =1$ the adjacency matrix $\mathbf{G}$ of the graph $\mathcal{G}_{f}(t,k,r)$ is $\mathbf{G}=$
	$$
	\left[\begin{array}{cc:cc|cc:cc|cc:cc|cc:cc}
		0 &  1 &  1 &  1 &  1 &  1 &  1 &  0 &  1 &  1 &  1 &  0 &  1 &  0 &  0 &  0 \\
		1 &  0 &  1 &  1 &  1 &  1 &  0 &  1 &  1 &  1 &  0 &  1 &  0 &  1 &  0 &  0 \\
		\hdashline
		1 &  1 &  0 &  1 &  1 &  0 &  1 &  1 &  1 &  0 &  1 &  1 &  0 &  0 &  1 &  0 \\
		1 &  1 &  1 &  0 &  0 &  1 &  1 &  1 &  0 &  1 &  1 &  1 &  0 &  0 &  0 &  1 \\
		\hline
		1 &  1 &  1 &  0 &  0 &  1 &  1 &  1 &  1 &  0 &  0 &  0 &  1 &  1 &  1 &  0 \\
		1 &  1 &  0 &  1 &  1 &  0 &  1 &  1 &  0 &  1 &  0 &  0 &  1 &  1 &  0 &  1 \\
		\hdashline
		1 &  0 &  1 &  1 &  1 &  1 &  0 &  1 &  0 &  0 &  1 &  0 &  1 &  0 &  1 &  1 \\
		0 &  1 &  1 &  1 &  1 &  1 &  1 &  0 &  0 &  0 &  0 &  1 &  0 &  1 &  1 &  1 \\
		\hline
		1 &  1 &  1 &  0 &  1 &  0 &  0 &  0 &  0 &  1 &  1 &  1 &  1 &  1 &  1 &  0 \\
		1 &  1 &  0 &  1 &  0 &  1 &  0 &  0 &  1 &  0 &  1 &  1 &  1 &  1 &  0 &  1 \\
		\hdashline
		1 &  0 &  1 &  1 &  0 &  0 &  1 &  0 &  1 &  1 &  0 &  1 &  1 &  0 &  1 &  1 \\
		0 &  1 &  1 &  1 &  0 &  0 &  0 &  1 &  1 &  1 &  1 &  0 &  0 &  1 &  1 &  1 \\
		\hline
		1 &  0 &  0 &  0 &  1 &  1 &  1 &  0 &  1 &  1 &  1 &  0 &  0 &  1 &  1 &  1 \\
		0 &  1 &  0 &  0 &  1 &  1 &  0 &  1 &  1 &  1 &  0 &  1 &  1 &  0 &  1 &  1 \\
		\hdashline
		0 &  0 &  1 &  0 &  1 &  0 &  1 &  1 &  1 &  0 &  1 &  1 &  1 &  1 &  0 &  1 \\
		0 &  0 &  0 &  1 &  0 &  1 &  1 &  1 &  0 &  1 &  1 &  1 &  1 &  1 &  1 &  0
	\end{array}\right],
	$${where the rows and columns of $\mathbf{G}$ are indexed by the vectors in $\mathbf{F}_2^4$ in lexicographic ordering.} The structure specified in Corollary \ref{corr: power of 2} can be seen in the above adjacency matrix.
\end{exmp}
\subsection{A lower bound on $r_{f}(k,t)$}

We now characterise the spectrum of the adjacency matrix of $\mathcal{G}_f(t,k,r)$.

\begin{thm}
	\label{hadamard diag}
	A $q^{n} \times q^{n}$ matrix $\mathbf{M}$ satisfying condition $C1$ is diagonalised by $\mathbf{W}_{q}^{\otimes n}$, where $\mathbf{W}_{q}$ is the $q$th order DFT matrix.
\end{thm}
\begin{IEEEproof}
	Let $\mathcal{M}_{n} \subseteq \mathbb{C}^{q^{n}\times q^{n}}$ be the set of $q^{n}\times q^{n}$ matrices satisfying condition $C1$. We can use an induction based argument to prove the claim. We can see that $\mathcal{M}_1$ is the set of all circulant matrices of order $q$. The DFT matrix $\mathbf{W}_q$ diagonalises all the matrices in $\mathcal{M}_1.$ Now, consider the induction hypothesis: $\mathbf{W}_q^{\otimes n}$ diagonalises all the matrices in $\mathcal{M}_n$. We have to show that $\mathbf{W}_q^{\otimes n+1}$ diagonalises all $\mathbf{M}\in \mathcal{M}_{n+1}$.
	
	Define the matrix $\mathbf{J}_i\in\mathcal{M}_1$ as $[\mathbf{J}_i]_{1k} = 1$ if $k=i$ and $0$ otherwise, i.e., $\mathbf{J}_i$ is the $q\times q$ circulant matrix with its first row equal to $\mathbf{e}_i$. Consider any $\mathbf{M} \in \mathcal{M}_{n+1}$, which is of the form
	
	$$
	\mathbf{M} = \begin{bmatrix}
		\mathbf{M}_{1} & \mathbf{M}_{2}&\mathbf{M}_3& \hdots &\mathbf{M}_{q-1}&\mathbf{M}_q\\
		
		\mathbf{M}_{q} & \mathbf{M}_{1}&\mathbf{M}_{2}& \hdots  &\mathbf{M}_{q-2}&\mathbf{M}_{q-1}\\
		
		\mathbf{M}_{q-1} & \mathbf{M}_{q}&\mathbf{M}_{1}& \hdots  &\mathbf{M}_{q-3}&\mathbf{M}_{q-2}\\
		\vdots & \vdots&\vdots& \ddots &\vdots& \vdots\\
		\mathbf{M}_{3} & \mathbf{M}_{4}&\mathbf{M}_{5}& \hdots  &\mathbf{M}_{1}&\mathbf{M}_{2}\\
		\mathbf{M}_2 & \mathbf{M}_{3}&\mathbf{M}_{4}& \hdots  &\mathbf{M}_{q}&\mathbf{M}_{1}\\
	\end{bmatrix}.
	$$
	Using Remark \ref{rem:rec circ}, we can express $\mathbf{M}$ as $\mathbf{M}= \sum_{i=1}^{q}\mathbf{J}_i\otimes\mathbf{M}_i$, where $\mathbf{M}_i\in \mathcal{M}_n\forall i\in[q]$. Then,
	\small\begin{align*}
		(\mathbf{W}_{q}^{\otimes n+1})^{\dagger}\mathbf{M}(\mathbf{W}_{q}^{\otimes n+1}) &= \sum_{i=1}^{q}(\mathbf{W}_{q}^{\otimes n+1})^\dagger(\mathbf{J}_i\otimes\mathbf{M}_i)(\mathbf{W}_{q}^{\otimes n+1})\\
		&=\sum_{i=1}^{q}(\mathbf{W}_q^\dagger\mathbf{J}_i\mathbf{W}_q)\otimes((\mathbf{W}_q^{\otimes n})^{\dagger}\mathbf{M}_i\mathbf{W}_q^{\otimes n}).
	\end{align*}
	\normalsize
	By the induction hypothesis, the above sum is a diagonal matrix. Thus, $\mathbf{W}_q^{\otimes n+1}$ diagonalises all $\mathbf{M}\in \mathcal{M}_{n+1}$. Hence, proved.
\end{IEEEproof}
Recall that the Hadamard matrix $\mathbf{H}_2 = \mathbf{W}_2$ and $\mathbf{H}_{2^n} = \mathbf{H}_2^{\otimes n}$. Thus, Corollary \ref{corr: power of 2} and Theorem \ref{hadamard diag} lead to the following result.

\begin{corollary}
	\label{corr:spectrum}
	
	For a linear function $f:\mathbb{F}_{2^m}^{k}\to\mathbb{F}_{2^m}^{l}$, the adjacency matrix $\mathbf{G}$ of the graph $\mathcal{G}_{f}(t,k,r)$ is a $2^{mn}\times 2^{mn}$ matrix that has the columns of $\mathbf{H}_{2^{mn}}$as its eigen vectors.
\end{corollary}
The spectrum of $\mathbf{G}$  can be used to get {bounds on the graph's independence number} as shown in \cite{cvetkovic}.

\begin{thm}
	\label{thm:cvetkovic}
	\cite{cvetkovic} For a linear function : $\mathbb{F}_q^k \to \mathbb{F}_q^l$, the independence number of the graph $\mathcal{G}_{f}(t,k,r)$, 
	$$\alpha \le \frac{-q^{k+r}\lambda_{min}(r)}{\lambda_{max}(r)-\lambda_{min}(r)},$$
	 where $\lambda_{max}(r)$ and $\lambda_{min}(r)$ denote the maximum and minimum eigen values respectively of the adjacency matrix of $\mathcal{G}_f(t,k,r)$.
\end{thm}
Note that for any graph, the least eigen value is non-positive and is $0$ only when there are no edges in the graph.

Theorem \ref{thm:cvetkovic}, Corollary \ref{corr:spectrum} and the fact that we need an independent set of size $q^k$, can be used to get a lower bound on $r_{f}(k,t)$ as shown in the following corollary.
\begin{corollary}
	\label{cor:eigen val bound}
	The redundancy $r$ of an $(f,t)$-FCC satisfies the inequality 
	$$q^r \ge 1-\frac{\lambda_{max}(r)}{\lambda_{min}(r)},$$
	 where $\lambda_{max}(r)$ and $\lambda_{min}(r)$ denote the maximum and minimum eigen values respectively of the adjacency matrix of $\mathcal{G}_f(t,k,r)$.
\end{corollary}

\begin{rem}
	The smallest $r$ that satisfies the inequality in Corollary \ref{cor:eigen val bound}, is a lower bound on the minimum redundancy $r_f(k,t)$.
\end{rem}

\begin{exmp}
	Consider the linear function  $f: \mathbb{F}_{2}^{3} \to \mathbb{F}_{2}^{2}$, defined as $f(\mathbf{u}) = \begin{bmatrix}
		0&1&1\\
		1&1&0
	\end{bmatrix}\mathbf{u}$. For $t=1\text{ and }r = 1$, it can be computed that $1-\frac{\lambda_{max}(1)}{\lambda_{min}(1)} = 6 > 2^1$. Similarly, for $t=1 \text{ and } r=2$, $1-\frac{\lambda_{max}(2)}{\lambda_{min}(2)} = 6 > 2^2$. Therefore, from Corollary \ref{cor:eigen val bound}, $(f,1)$-FCCs of parity length $1$ or $2$ do not exist. For $t=1 \text{ and } r=3$, $1-\frac{\lambda_{max}(3)}{\lambda_{min}(3)} = 6.5 < 2^3$. Thus, from Corollary \ref{cor:eigen val bound}, the optimal redundancy $r_f(3,1) \ge 3$.
\end{exmp}	
\subsection{A class of linear functions for which coset-wise coding leads to optimal FCCs}
\label{subsec: thm 5 tight}

In coset-wise coding all the message vectors in a particular coset in $\mathbb{F}_{q}^{k}/ker(f)$ are given the same parity vector. Such a scheme was proposed in \cite{FCC}, as it reduces the complexity of encoding. As shown in Theorem \ref{thm : df UB}, for such a coding scheme, the distance constraint
is specified by the FDM, $\mathbf{D}_{f}(t,f_{0}, f_{1} \hdots f_{q^{l}-1})$. For a linear function $f : \mathbb{F}_{q}^{k} \to \mathbb{F}_{q}^{l}$, there is a reduction in the dimension
of the function correction problem from $k$ to $l$ while doing coset-wise coding.

Now, we look at a class of functions for which the optimal redundancy $r_f(k,t) = N_q(\mathbf{D}_{f}(t,f_{0}, f_{1} \hdots f_{|Im(f)|-1}))$. From Remark \ref{rem:coset leaders} and Corollary \ref{label:optimal coding}, we can see that for linear functions, the upper bound in Theorem \ref{thm : df UB} is tight, if there is a selection of minimum weight representative vectors from each coset in $\mathbb{F}_{q}^{k}/ker(f)$ such that it satisfies (\ref{rep vectors}). Now, we give a class of functions for which such a selection of minimum weight vectors is possible.

\begin{defn}
	\label{def:rep vectors}
	For a linear function $f:\mathbb{F}_{q}^{k} \to \mathbb{F}_{q}^{l}$ define $$ \mathcal{S} \coloneqq \{arg\min_{\mathbf{u}\in \bar{f}}w_{H}(\mathbf{u}) : \bar{f} \in \mathbb{F}_{q}^{k}/ker(f)\},$$
	i.e., $\mathcal{S}$ is a selection of minimum weight vectors from each coset in $\mathbb{F}_{q}^{k}/ker(f)$. Note that $\mathcal{S}$ is not unique.
	
\end{defn}

\begin{defn}
	\label{no of cosets}
	We define $C_{f}(i)$ to be the number of cosets in $\mathbb{F}_{q}^{k}/ker(f)$ whose minimum weight vector has Hamming weight $i$ with $1s$ in $i$ indices, $i=1,2,\cdots k,$ and $0s$ in all the other indices, i.e.,
	\begin{equation*}
		C_{f}(i)= \left|\{\mathbf{u}\in \mathcal{S}: \mathbf{u} = \sum_{j\in I}^{}\mathbf{e}_j \text{ for some } I \subseteq [k], |I|=i\}\right|.
	\end{equation*}
\end{defn}

\begin{lem}
	\label{l coset}
	For a linear function $f(\mathbf{u}) = \mathbf{F}\mathbf{u}$ defined on  $\mathbb{F}_{q}^{k}$ 
	at least $rank_{\mathbb{F}_q}(\mathbf{F})$ cosets in $\mathbb{F}_{q}^{k}/ker(f)$ should have a unit vector as its minimum weight vector, i.e., $C_f(1)\ge rank_{\mathbb{F}_q}(\mathbf{F})$. 
\end{lem}
\begin{IEEEproof}
	Consider the linear function $f(\mathbf{u}) = \mathbf{F}\mathbf{u}$. We have, $C_{f}(1)= \left|\{\mathbf{u}\in \mathcal{S}: \mathbf{u} = \mathbf{e}_j \text{ for some } j\in [k]\}\right|.$
	The function $f$ maps all the vectors in a particular coset $\bar{f_{i}}\in \mathbb{F}_{q}^{k}/ker(f)$ to the same image $f_{i} \in \mathbb{F}_{q}^{l}$ and vectors in distinct cosets will not have the same image. This implies that for some $i,j \in [k]$, unit vectors $\mathbf{e}_{i}$ and $\mathbf{e}_{j}$ belong to the same coset if and only if the $i-$th and the $j-$th columns, $\mathbf{F}_{(i)}\text{ and }\mathbf{F}_{(j)}$ are equal. This implies that the number of distinct non-zero columns of $\mathbf{F}$ is equal to $C_{f}(1)$. Since $rank_{\mathbb{F}_q}(\mathbf{F})$ cannot be more than the number of distinct non-zero columns of $\mathbf{F}$, $C_{f}(1) \ge rank_{\mathbb{F}_q}(\mathbf{F})$.
\end{IEEEproof}

\begin{lem}
	\label{vec space coset}
	Consider a linear function $f:\mathbb{F}_{q}^{k} \to \mathbb{F}_{q}^{l}$. There exists a selection $\mathcal{S} \coloneqq \{arg\min_{\mathbf{u}\in \bar{f}}w_{H}(\mathbf{u}) : \bar{f} \in \mathbb{F}_{q}^{k}/ker(f)\}$ that forms a subspace of $\mathbb{F}_{q}^{k}$ if and only if exactly $l$ cosets of $\mathbb{F}_{q}^{k}/ker(f)$ have a unit vector as its minimum weight vector, i.e., $C_f(1) = l$.
\end{lem}
\begin{IEEEproof}
	For the forward implication, consider a selection $\mathcal{S}$ which is a subspace of $\mathbb{F}_{q}^{k}$. Note that $dim_{q}\mathcal{S}=l$. 
	From Lemma \ref{l coset}, $C_{f}(1) = l'\ge l$. 
	For the sake of contradiction, let $l'> l$. Then, there exists a set of $l'$ unit vectors in $\mathcal{S}$, which are independent. This is a contradiction as $dim_{q}\mathcal{S} = l$.
	
	For the reverse implication, let $C_{f}(1) = l$. Let $E = \{\mathbf{u}\in \mathcal{S}: \mathbf{u} = \mathbf{e}_j \text{ for some } j\in [k]\}$. We now show that vectors in $span(E)$ occur in distinct cosets.
	Assume WLOG that $E = \{\mathbf{e}_{1}, \mathbf{e}_{2}, \mathbf{e}_{3},...,\mathbf{e}_{l}\}$. Since $E$ is a set of independent vectors, $\mathbf{F}E \coloneq \{\mathbf{F}\mathbf{e}_{1}, \mathbf{F}\mathbf{e}_{2}, \mathbf{F}\mathbf{e}_{3},..., \mathbf{F}\mathbf{e}_{l}\}$ is a basis of $\mathbb{F}_{q}^{l}$, the image of $f$.
	
	Consider $\mathbf{v}_{1}, \mathbf{v}_{2} \in span(E)$. Let $\mathbf{v}_{1} = \sum_{i=1}^{l}a_{i}\mathbf{e}_{i}$ and $\mathbf{v}_{2} = \sum_{i=1}^{l}b_{i}\mathbf{e}_{i}$, where $a_i,b_i \in \mathbb{F}_q \forall i \in [l].$ Then,
	\begin{align*}
		&\mathbf{v}_{1} \text{ and } \mathbf{v}_{2} \text{ are in the same coset}\\
		&\iff \mathbf{v}_{1}-\mathbf{v}_{2} \in ker(f)\\
		&\iff \mathbf{F}(\mathbf{v}_{1}-\mathbf{v}_{2}) = 0\\
		&\iff \mathbf{F}(\sum_{i=1}^{l}a_{i}\mathbf{e}_{i}-\sum_{i=1}^{l}b_{i}\mathbf{e}_{i}) = 0\\
		&\iff \sum_{i=1}^{l}a_{i}\mathbf{F}\mathbf{e}_{i}-\sum_{i=1}^{l}b_{i}\mathbf{F}\mathbf{e}_{i} = 0\\
		&\iff a_{i}=b_{i} \text{ } \forall \text{ } i \in [l]\text{ }\{\because \mathbf{F}E \text{ is a basis of } \mathbb{F}_{q}^{l}\}.
	\end{align*} 
	
	So, distinct vectors in $span(E)$ are present in distinct cosets. It is easy to verify that the vectors in $span(E)$ are minimum weight vectors in the corresponding cosets. Thus, $\mathcal{S} = span(E)$.
\end{IEEEproof}

Thus, for a selection $\mathcal{S}$ which is a vector sub-space to exist, exactly $l$ cosets should have unit vectors as minimum weight vectors. So, the matrix representation $\mathbf{F}$ of $f$ must have exactly $l$ distinct non-zero columns. Since $rank_{\mathbb{F}_q}(\mathbf{F})=l$, these columns have to be independent.

Lemmas \ref{l coset} and \ref{vec space coset} lead to a class of functions for which the representative vectors satisfy (\ref{rep vectors}) which is given in the following theorem.
\begin{thm}
	\label{class A}
	Consider a linear function 
	\begin{equation*}
		f : \mathbb{F}_{q}^{k} \to \mathbb{F}_{q}^{l}
	\end{equation*}
	\begin{equation*}
		\mathbf{u}  \mapsto \mathbf{F}\mathbf{u};~~~ \mathbf{F} \in \mathbb{F}_{q}^{l \times k}\text{ and } rank(\mathbf{F})=l.
	\end{equation*}
	Then, there exists a selection of representative vectors $\{\mathbf{u}_{0}, \mathbf{u}_{1} \hdots \mathbf{u}_{q^{l}-1}\}$, where $\{f(\mathbf{u}_{0}), f(\mathbf{u}_{1}) \hdots f(\mathbf{u}_{q^{l}-1})\} = Im(f)$ such that $\mathbf{D}_{f}(t,f_{0}, f_{1} \hdots f_{q^{l}-1}) = \mathbf{D}_{f}(t,\mathbf{u}_{0}, \mathbf{u}_{1} \hdots \mathbf{u}_{q^{l}-1})$ if the number of   distinct non-zero columns of $\mathbf{F}$ is exactly $l$.
\end{thm}

\begin{IEEEproof}
	From Remark \ref{rem:coset leaders}, we know that for linear functions, the weight distribution of the vectors in $\mathcal{S}=\{\mathbf{u}_{0}, \mathbf{u}_{1} \hdots \mathbf{u}_{q^{l}-1}\}$ is the same as $\{d_f(f_{i},0): f_{i}\in \mathbb{F}_{q}^{l}\}$, the intercoset distance distribution. According to Lemma \ref{vec space coset}, there should be exactly $l$ unit vectors in $\mathcal{S}$ for it to be a vector space and this implies that $\mathbf{F}$ should have exactly $l$ non-zero distinct columns. Since $rank_{\mathbb{F}_q}(\mathbf{F})$ is assumed to be $l$, the non-zero columns are also independent. If $\mathcal{S}$ is a vector sub-space, then its distance distribution will also be uniform, which implies that $\mathbf{D}_{f}(t,f_{0}, f_{1} \hdots f_{q^{l}-1}) = \mathbf{D}_{f}(t,\mathbf{u}_{0}, \mathbf{u}_{1} \hdots \mathbf{u}_{q^{l}-1})$.

\end{IEEEproof}

The function $f$ given in Example \ref{exmp:explanation} has exactly 2 distinct, non-zero columns independent over $\mathbb{F}_{2}$. We can see that there exists a selection $\mathcal{S} = \{0000,0100,1000,1100\}$ which is a subspace of $\mathbb{F}_{2}^{4}$ and $\mathbf{D}_{f}(t,f_{0}, f_{1}, f_{2}, f_{3}) = \mathbf{D}_{f}(t,\mathbf{u}_{0}, \mathbf{u}_{1}, \mathbf{u}_{2}, \mathbf{u}_{3})$, where $\mathcal{S}=\{\mathbf{u}_{0}, \mathbf{u}_{1}, \mathbf{u}_{2}, \mathbf{u}_{3}\}$. 

\subsection{A coding scheme for the class of functions specified in Theorem \ref{class A}}
\label{subsec: thm 5 coding scheme}

We now provide a coding scheme for the class of functions specified in Theorem \ref{class A}. Consider a function $f : \mathbb{F}_{q}^{k} \to \mathbb{F}_{q}^{l}$ with $\mathbf{F} \in \mathbb{F}_{q}^{l \times k}$ as its matrix representation which has exactly $l$ distinct non-zero independent columns. Choose the set of minimum weight representative vectors $ \mathcal{S} =\{\mathbf{u}_{0}, \mathbf{u}_{1} \hdots \mathbf{u}_{q^{l}-1}\}$ such that $\mathcal{S}$ is a subspace of $\mathbb{F}_{q}^{k}$. 
From Corollary \ref{label:optimal coding}, the optimal redundancy is given by $r_{f}(k,t) = N_{q}(\mathbf{D}_{f}(t,f_{0}, f_{1} \hdots f_{q^{l}-1})) = N_{q}( \mathbf{D}_{f}(t,\mathbf{u}_{0}, \mathbf{u}_{1} \hdots \mathbf{u}_{q^{l}-1}))$. Let $E$ be a spanning set of $\mathcal{S}$. Assume WLOG that $E=\{\mathbf{e}_{1}, \mathbf{e}_{2}, \hdots \mathbf{e}_{l}\}$. Then $\mathcal{S}$ is of the following form:
$$
\mathcal{S} = \{(x_{1} x_{2} \hdots x_{k}) \in \mathbb{F}_{q}^{k}: x_{i} = 0 \text{ }\forall \text{ }i \in [l+1:k]\}.
$$
Define,
$$
\mathcal{S}' = \{(x_{1} x_{2} \hdots x_{l}): (x_{1} x_{2} \hdots x_{l} \hdots x_{k}) \in \mathcal{S}\}.
$$
Note that $\mathcal{S}' \equiv \mathbb{F}_{q}^{l}$. 
Let $\mathcal{S}' = \{\mathbf{u}'_{0}, \mathbf{u}'_{1} \hdots \mathbf{u}'_{q^{l}-1}\}$. Note that $\mathbf{D}_{f}(t,\mathbf{u}'_{0}, \mathbf{u}'_{1} \hdots \mathbf{u}'_{q^{l}-1}) = \mathbf{D}_{f}(t,\mathbf{u}_{0}, \mathbf{u}_{1} \hdots \mathbf{u}_{q^{l}-1})$. 

Note that the DRM  $\mathbf{D}_{f}(t,\mathbf{u}'_{0}, \mathbf{u}'_{1} \hdots \mathbf{u}'_{q^{l}-1})$ is the same as the DRM for the parity vectors of a classical systematic ECC of cardinality $q^{l}$ and minimum distance $2t+1$ (The parity vectors of a classical systematic ECC can be considered as the parity vectors of an FCC where the function to be corrected is a bijection). So, the parity bits of a systematic $(n,q^{l},2t+1)$ block code will be a $\mathbf{D}$-code where $\mathbf{D} = \mathbf{D}_{f}(t,\mathbf{u}_{0}, \mathbf{u}_{1} \hdots \mathbf{u}_{q^{l}-1})$.

\begin{rem}
	The coding scheme given above gives optimal constructions of FCCs for certain parameters. Consider a function $f : \mathbb{F}_{q}^{k} \to \mathbb{F}_{q}^{l}$ with $\mathbf{F} \in \mathbb{F}_{q}^{l \times k}$ as its matrix representation which has exactly $l$ distinct non-zero independent columns. If an optimal classical error correcting of cardinality $q^l$ and minimum distance $2t+1$ exists, an optimal $(f,t)$-FCC can be constructed using the above coding scheme, as shown in Examples \ref{exmp:opt 1} and \ref{exmp:opt 2}.
\end{rem}

\begin{exmp}
	\label{exmp:opt 1}
	For the function in Example \ref{exmp:explanation}, the matrix $\mathbf{F} = \begin{bmatrix}
		1&1&1&0\\
		0&1&1&0
	\end{bmatrix}$ satisfies the above condition. We can choose $\mathcal{S} = \{0000,1000,0100,1100\}$ and $\mathcal{S}'=\{00,01,10,11\} \equiv\mathbb{F}_{2}^{2}$. The FDM  is the same as $\mathbf{D}_{f}(t,00,01,10,11)$. Say, we need an $(f,1)$-FCC. This is equivalent to the distance constraint that should be satisfied by the parity bits of a binary block code of cardinality 4 and minimum distance 3. From \cite{codetables}, there exists an optimal $(5,4,3)_2$ systematic block code. Using the 3 parity bits of this code for coset-wise coding, we have an FCC of $r=3$. Thus, for the function in Example \ref{exmp:explanation}, $r_{f}(4,1) = 3$.
\end{exmp}

\begin{exmp}
	\label{exmp:opt 2}
	Consider the function $f: \mathbb{F}_3^3 \rightarrow \mathbb{F}_3^2$ defined as $f(\mathbf{u})=\left[\begin{array}{lll}2 & 2 & 0 \\ 1 & 1 & 1\end{array}\right] \mathbf{u}$. The kernel, $ker(f)=$ $\{000,120,210\}$. The coset decomposition $\mathbb{F}_3^3 / ker(f)$ is
	
	$$
	\begin{gathered}
		\{\{000,120,210\}, ~~~
		\{001,121,211\}, \\
		\{010,100,220\}, ~~~
		\{002,122,212\}, \\
		\{020,110,200\}, ~~~
		\{011,101,221\}, \\
		\{111,201,021\}, ~~~
		\{222,012,102\}, \\
		\{112,202,022\}\}.
	\end{gathered}
	$$
	The matrix $\left[\begin{array}{lll}2 & 2 & 0 \\ 1 & 1 & 1\end{array}\right]$ has $2$ distinct non-zero columns. Thus, we have a selection $$\mathcal{S} = \{000,001,010,002,020,011,022,012,021\},$$ which is a subspace of $\mathbb{F}_3^2$ and $\mathbf{D}_{f}(t,f_{0}, f_{1}, \hdots, f_{8}) = \mathbf{D}_{f}(t,\mathbf{u}_{0}, \mathbf{u}_{1}, \hdots, \mathbf{u}_{8})$, where $\mathcal{S} =\{\mathbf{u}_{0}, \mathbf{u}_{1}, \mathbf{u}_{2}, \hdots, \mathbf{u}_{8}\}$. As mentioned in the encoding scheme, we can now define $\mathcal{S}' \coloneq \{\mathbf{u}'_0,\mathbf{u}'_1, \hdots, \mathbf{u}'_8\} = \{00,01,10,02,20,11,22,12,21\}$. Say, we need to construct an $(f,1)$-FCC. This distance requirement is equivalent to the distance requirement that should be satisfied by the parity bits of a ternary block code of cardinality $9$ and minimum distance 3. From \cite{codetables}, there exists an optimal $(4,9,3)_3$ systematic block code. Using the $2$ parity bits of this code for coset-wise coding, we have $r_f(k,1) = 2$.
\end{exmp}

\subsection{A class of linear functions for which coset-wise coding is equivalent to a reduced dimensional classical error correction problem.}
\label{subsec: eq to classical}

Now, we consider linear functions for which coset-wise coding is equivalent to a reduced dimensional classical error correction problem. Consider a linear function,  
$
f : \mathbb{F}_{q}^{k} \to \mathbb{F}_{q}^{l}
$ and
$
\mathbf{u}  \mapsto \mathbf{F}\mathbf{u}; \mathbf{F} \in \mathbb{F}_{q}^{l \times k}
$
with $k\ge q^{l}-1$. If the number of distinct columns of $\mathbf{F}$ is at least $q^{l}-1$, then all the cosets of $\mathbb{F}_{q}^{k}/ker(f)$ will have a unit vector. Then, the FDM will be of the form 
\begin{equation}
	[\mathbf{D}_{f}(t,f_{0}, f_{1} \hdots f_{q^{l}-1})]_{ij} =
	\begin{cases}
		
		2t;&\text{if }i \ne j\\
		0; &  else,
	\end{cases} 
\end{equation}
i.e, the diagonal entries are $0$ and the non-diagonal entries are $2t$.

So, the parity vectors satisfying the distance requirements will be an ECC of cardinality $q^{l}$ and minimum distance $2t$. Essentially we are reducing the dimension of the error correction problem from $k \ge q^{l}$ to $l$ and minimum distance requirement from $2t+1$ to $2t$.

\section{Discussion}
We proposed a lower bound for the redundancy required for an $(f,t)$-FCC with $2t+1 \le k$, which also led to a bound on the redundancy of classical ECCs with minimum distance $d\le k$. We showed that MDS codes whose minimum distance is less than their dimension are a class of codes whose redundancy satisfies our bound with equality. We compared our bound with the ZLL and BGS bound. We emphasised the use of the graphical representation of FCCs proposed in \cite{FCC} to study systematic ECCs. Our approach of representing $\mathcal{G}_{f}(t,k,r)$ as a Cartesian product makes the obtained bound non-trivial only in the region $2t+1 \le k$. Better bounds on the independence number of the graph $\mathcal{G}_{f}(t,k,r)$ will give better bounds on the redundancy of FCCs and ECCs

We explored a class of function-correcting codes where the function to be computed at the receiver is linear. We showed that the adjacency matrix of $\mathcal{G}_{f}(t,k,r)$ has a recursive block circulant structure. When the domain of the function is a vector space over a field of characteristic 2, we characterised the spectrum of the adjacency matrix, which leads to lower bounds on redundancy. Functions that can attain this lower bound are yet to be characterised. This bound can also be tailored for systematic ECCs, but the bound does not give a lot of information as there is no closed form expression for the eigen values of the graph. The structure induced on the graph by linear bijections is yet to be studied for a better characterisation of the eigen values. For linear functions, we showed that the generalised Plotkin bound proposed in \cite{FCC} simplifies. We characterised a class of functions for which coset-wise coding is optimal and proposed a coding scheme for such functions. We also characterised a class of functions for which the redundancy to be added for coset-wise coding is equivalent to a lower dimensional classical ECC. Further research directions include using the structure of $\mathcal{G}_{f}(t,k,r)$ to get improved bounds and coding schemes and to identify functions and coding schemes that can attain or come close to the proposed bounds. Linear codes for function correction are yet to be explored.

\begin{IEEEbiographynophoto}{Rohit Premlal}
	(Graduate Student Member, IEEE) received the B.Tech. degree in electronics and communication engineering from the College of Engineering, Trivandrum, in 2021. He received the M.Tech degree in Electronics and Communication Engineering from the Indian Institute of Science, Bangalore, in 2024. He is currently pursuing the Ph.D. degree with the Department of Electrical and Computer Engineering, University of Maryland, College Park. 
\end{IEEEbiographynophoto}

\begin{IEEEbiographynophoto}{B. Sundar Rajan}
	(Life Fellow, IEEE) was born in Tamil Nadu, India. He received the B.Sc. degree
	in mathematics from Madras University, Madras, India, in 1979, the B.Tech. degree in electronics from the Madras Institute of Technology, Madras, in 1982, and the M.Tech. and Ph.D. degrees in electrical engineering from IIT Kanpur, Kanpur, in 1984 and 1989, respectively.
	
	He was a Faculty Member at the Department of Electrical Engineering, IIT Delhi, New Delhi, from 1990 to 1997. He has been a Professor with the Department of Electrical Communication Engineering, Indian Institute of Science, Bengaluru, since 1998. His primary research interests include space-time coding for MIMO channels, distributed space-time coding and cooperative communication, coding for multiple-access and relay channels, and network coding.
	
	Dr. Rajan is a J. C. Bose National Fellow (2016–2025) and a member of the American Mathematical Society. He is a fellow of the Indian National Academy of Engineering, the Indian National Science Academy, the Indian Academy of Sciences, and the National Academy of Sciences, India. He was a recipient of Prof. Rustum Choksi Award by IISc for Excellence in Research in Engineering in 2009, the IETE Pune Center’s S. V. C. Aiya Award for Telecom Education in 2004, and the Best Academic Paper Award at IEEE WCNC in 2011. He served as the Technical Program Co-Chair for IEEE Information Theory Workshop (ITW’02) held in Bengaluru, in 2002. He was an Editor of IEEE \textsc{Wireless Communications Letters} (2012–2015) and IEEE \textsc{Transactions on Wireless Communications} (2007–2011), and an Associate Editor of Coding Theory for IEEE \textsc{Transactions on Information Theory} (2008–2011 and 2013–2015). 
\end{IEEEbiographynophoto}

\end{document}